\documentclass[conference,a4paper]{IEEEtran}

\usepackage{mdwmath}
\usepackage{mdwtab}
\usepackage{tikz}

\usepackage{graphicx}
\usepackage{amsmath}
\usepackage{latexsym}
\usepackage{amssymb}
\usepackage{comment}
\usepackage{cite}
\usepackage{url}
\usepackage[T3,T1]{fontenc}
\DeclareSymbolFont{tipa}{T3}{cmr}{m}{n}
\DeclareMathAccent{\invbreve}{\mathalpha}{tipa}{16}

\global\long\def\L{\mathsf{L}}
\global\long\def\D{\mathsf{D}}
\global\long\def\Sgen{\mathcal{S}_\mathsf{gen}}
\global\long\def\Kgen{\mathcal{K}_\mathsf{gen}}

\begin{document}
\title{
New Distributed Source Encryption Framework 
%FUsing Correlated Keys 
}
\author{%
	\IEEEauthorblockN{Yasutada Oohama and Bagus Santoso}
	\IEEEauthorblockA{University of Electro-Communications, Tokyo, Japan\\ 
	Email: \url{{oohama,santoso.bagus}@uec.ac.jp}}
%	\and 
%	\IEEEauthorblockN{Yasutada Oohama}
%	\IEEEauthorblockA{University of Electro-Communications\\ 
%
}
\maketitle

\begin{abstract}
We pose and investigate the distributed secure source 
coding based on the common key cryptosystem. 
This cryptosystem includes 
the secrecy amplification problem for 
distributed encrypted sources 
with correlated keys using post-encryption-compression, which 
was posed investigated by Santoso and Oohama. 
In this paper we propose a new security criterion which 
is more natural compared with the commonly used security 
criterion which is based on the upper-bound of mutual information 
between the plaintext and the ciphertext. Under this criterion, 
we establish the necessary and sufficient condition for 
the secure transmission of correlated sources.
\end{abstract}
%Our result yields that    
%(110%(\cite{}. 
%We study the secure source coing based on the common 
%We study side-channel attacks for the Shannon cipher system. 
%To pose side channel-attacks to the Shannon cipher system, 
%we regard them as a signal estimation via encoded data from 
%two distributed sensors. This can be formulated as the one helper 
%source coding problem posed and investigated
%by Ahlswede, K\"orner(1975), and Wyner(1975). We further investigate 
%the posed problem to derive new secrecy bounds. Our results 
%are derived by a coupling of the result Watanabe and Oohama(2012) 
%obtained on bounded storage eavesdropper with the exponential strong converse 
%theorem Oohama(2015) established for the one helper 
%source coding problem.
%\end{abstract}

% \fi
\IEEEpeerreviewmaketitle

%%%%%%%%QED%%%%%%%%%%%%
\newcommand{\qed}{\hfill$\square$}
\newcommand{\suchthat}{\mbox{~s.t.~}}
\newcommand{\markov}{\leftrightarrow}

%%%%%%% argmax argmin %%%%%%%%%%%%
\newcommand{\argmax}{\mathop{\rm argmax}\limits}
\newcommand{\argmin}{\mathop{\rm argmin}\limits}

\newcommand{\ExP}{\rm e}

\newcommand{\Cls}{class NL}
\newcommand{\vSpa}{\vspace{0.3mm}}
\newcommand{\Prmt}{\zeta}
\newcommand{\pj}{\omega_n}

\newfont{\bg}{cmr10 scaled \magstep4}
\newcommand{\bigzerol}{\smash{\hbox{\bg 0}}}
\newcommand{\bigzerou}{\smash{\lower1.7ex\hbox{\bg 0}}}
\newcommand{\nbn}{\frac{1}{n}}
\newcommand{\ra}{\rightarrow}
\newcommand{\la}{\leftarrow}
\newcommand{\ldo}{\ldots}
\newcommand{\typi}{A_{\epsilon }^{n}}
\newcommand{\bx}{\hspace*{\fill}$\Box$}
\newcommand{\pa}{\vert}
\newcommand{\ignore}[1]{}

%%%%%%%%%proof environment%%%%%%%%

\newtheorem{proposition}{Proposition}
\newtheorem{definition}{Definition}
\newtheorem{theorem}{Theorem}
\newtheorem{lemma}{Lemma}
\newtheorem{corollary}{Corollary}
\newtheorem{remark}{Remark}
\newtheorem{property}{Property}

\newcommand{\defeq}{:=}

\newcommand{\Qed}{\hbox{\rule[-2pt]{3pt}{6pt}}}
\newcommand{\beq}{\begin{equation}}
\newcommand{\eeq}{\end{equation}}
\newcommand{\beqa}{\begin{eqnarray}}
\newcommand{\eeqa}{\end{eqnarray}}
\newcommand{\beqno}{\begin{eqnarray*}}
\newcommand{\eeqno}{\end{eqnarray*}}
\newcommand{\ba}{\begin{array}}
\newcommand{\ea}{\end{array}}

\newcommand{\vc}[1]{\mbox{\boldmath $#1$}}
\newcommand{\lvc}[1]{\mbox{\scriptsize \boldmath $#1$}}
\newcommand{\svc}[1]{\mbox{\tiny \boldmath $#1$}}

\newcommand{\wh}{\widehat}
\newcommand{\wt}{\widetilde}
\newcommand{\ts}{\textstyle}
\newcommand{\ds}{\displaystyle}
\newcommand{\scs}{\scriptstyle}
\newcommand{\vep}{\varepsilon}
\newcommand{\rhp}{\rightharpoonup}
\newcommand{\cl}{\circ\!\!\!\!\!-}
\newcommand{\bcs}{\dot{\,}.\dot{\,}}
\newcommand{\eqv}{\Leftrightarrow}
\newcommand{\leqv}{\Longleftrightarrow}

\newcommand{\irr}[1]{{\color[named]{Red}#1\normalcolor}}

\newcommand{\hugel}{{\arraycolsep 0mm
                    \left\{\ba{l}{\,}\\{\,}\ea\right.\!\!}}
\newcommand{\Hugel}{{\arraycolsep 0mm
                    \left\{\ba{l}{\,}\\{\,}\\{\,}\ea\right.\!\!}}
\newcommand{\HUgel}{{\arraycolsep 0mm
                    \left\{\ba{l}{\,}\\{\,}\\{\,}\vspace{-1mm}
                    \\{\,}\ea\right.\!\!}}
\newcommand{\huger}{{\arraycolsep 0mm
                    \left.\ba{l}{\,}\\{\,}\ea\!\!\right\}}}
\newcommand{\Huger}{{\arraycolsep 0mm
                    \left.\ba{l}{\,}\\{\,}\\{\,}\ea\!\!\right\}}}
\newcommand{\HUger}{{\arraycolsep 0mm
                    \left.\ba{l}{\,}\\{\,}\\{\,}\vspace{-1mm}
                    \\{\,}\ea\!\!\right\}}}

\newcommand{\hugebl}{{\arraycolsep 0mm
                    \left[\ba{l}{\,}\\{\,}\ea\right.\!\!}}
\newcommand{\Hugebl}{{\arraycolsep 0mm
                    \left[\ba{l}{\,}\\{\,}\\{\,}\ea\right.\!\!}}
\newcommand{\HUgebl}{{\arraycolsep 0mm
                    \left[\ba{l}{\,}\\{\,}\\{\,}\vspace{-1mm}
                    \\{\,}\ea\right.\!\!}}
\newcommand{\hugebr}{{\arraycolsep 0mm
                    \left.\ba{l}{\,}\\{\,}\ea\!\!\right]}}
\newcommand{\Hugebr}{{\arraycolsep 0mm
                    \left.\ba{l}{\,}\\{\,}\\{\,}\ea\!\!\right]}}
\newcommand{\HUgebr}{{\arraycolsep 0mm
                    \left.\ba{l}{\,}\\{\,}\\{\,}\vspace{-1mm}
                    \\{\,}\ea\!\!\right]}}

\newcommand{\hugecl}{{\arraycolsep 0mm
                    \left(\ba{l}{\,}\\{\,}\ea\right.\!\!}}
\newcommand{\Hugecl}{{\arraycolsep 0mm
                    \left(\ba{l}{\,}\\{\,}\\{\,}\ea\right.\!\!}}
\newcommand{\hugecr}{{\arraycolsep 0mm
                    \left.\ba{l}{\,}\\{\,}\ea\!\!\right)}}
\newcommand{\Hugecr}{{\arraycolsep 0mm
                    \left.\ba{l}{\,}\\{\,}\\{\,}\ea\!\!\right)}}

\newcommand{\hugepl}{{\arraycolsep 0mm
                    \left|\ba{l}{\,}\\{\,}\ea\right.\!\!}}
\newcommand{\Hugepl}{{\arraycolsep 0mm
                    \left|\ba{l}{\,}\\{\,}\\{\,}\ea\right.\!\!}}
\newcommand{\hugepr}{{\arraycolsep 0mm
                    \left.\ba{l}{\,}\\{\,}\ea\!\!\right|}}
\newcommand{\Hugepr}{{\arraycolsep 0mm
                    \left.\ba{l}{\,}\\{\,}\\{\,}\ea\!\!\right|}}

\newcommand{\MEq}[1]{\stackrel{%\mbox
{\rm (#1)}}{=}}

\newcommand{\MLeq}[1]{\stackrel{%\mbox
{\rm (#1)}}{\leq}}

\newcommand{\ML}[1]{\stackrel{%\mbox
{\rm (#1)}}{<}}

\newcommand{\MGeq}[1]{\stackrel{%\mbox
{\rm (#1)}}{\geq}}

\newcommand{\MG}[1]{\stackrel{%\mbox
{\rm (#1)}}{>}}

\newcommand{\MPreq}[1]{\stackrel{%\mbox
{\rm (#1)}}{\preceq}}

\newcommand{\MSueq}[1]{\stackrel{%\mbox
{\rm (#1)}}{\succeq}}

\newcommand{\MSubeq}[1]{\stackrel{%\mbox
{\rm (#1)}}{\subseteq}}

\newcommand{\MSupeq}[1]{\stackrel{%\mbox
{\rm (#1)}}{\supseteq}}

\newcommand{\MRarrow}[1]{\stackrel{%\mbox
{\rm (#1)}}{\Rightarrow}}

\newcommand{\MLarrow}[1]{\stackrel{%\mbox
{\rm (#1)}}{\Leftarrow}}

\newcommand{\SZZpp}{%%%%%%%%%%%%%%%%%%%%%%%%%%%%%%%%%%%%%%

\newcommand{\vcc}{{c}^n}
\newcommand{\vck}{{k}^n}
\newcommand{\vcx}{{x}^n}
\newcommand{\vcy}{{y}^n}
\newcommand{\vcz}{{z}^n}
\newcommand{\vckone}{{k}_1^n}
\newcommand{\vcktwo}{{k}_2^n}
\newcommand{\vcxone}{{x}^n}

\newcommand{\vcxtwo}{{x}_2^n}
\newcommand{\vcyone}{{y}_1^n}
\newcommand{\vcytwo}{{y}_2^n}

\newcommand{\cvcx}{\check{x}^n}
\newcommand{\cvcy}{\check{y}^n}
\newcommand{\cvcz}{\check{z}^n}
\newcommand{\cvcxone}{\check{x}^n}

\newcommand{\cvcxtwo}{\check{x}_2^n}

\newcommand{\hvcx}{\widehat{x}^n}
\newcommand{\hvcy}{\widehat{y}^n}
\newcommand{\hvcz}{\widehat{z}^n}
\newcommand{\hvckone}{\widehat{k}_1^n}
\newcommand{\hvcktwo}{\widehat{k}_2^n}

\newcommand{\hvcxone}{\widehat{x}^n}

\newcommand{\hvcxtwo}{\widehat{x}_2^n}

\newcommand{\lvcc}{{c}^n}
\newcommand{\lvck}{{k}^n}
\newcommand{\lvcx}{{x}^n}
\newcommand{\lvcy}{{y}^n}
\newcommand{\lvcz}{{z}^n}

\newcommand{\lvckone}{{k}_1^n}
\newcommand{\lvcktwo}{{k}_2^n}
\newcommand{\lvcxone}{{x}^n}

\newcommand{\lvcxtwo}{{x}_2^n}
\newcommand{\lvcyone}{{y}_1^n}
\newcommand{\lvcytwo}{{y}_2^n}

\newcommand{\clvcxone}{\check{x}^n}

\newcommand{\clvcxtwo}{\check{x}_2^n}

\newcommand{\hlvckone}{\widehat{k}_1^n}
\newcommand{\hlvcktwo}{\widehat{k}_2^n}

\newcommand{\hlvcxone}{\widehat{x}^n}

\newcommand{\hlvcxtwo}{\widehat{x}_2^n}

\newcommand{\rvcc}{{C}^n}
\newcommand{\rvck}{{K}^n}
\newcommand{\rvcx}{{X}^n}
\newcommand{\rvcy}{{Y}^n}
\newcommand{\rvcz}{{Z}^n}
\newcommand{\rvccone}{{C}_1^n}
\newcommand{\rvcctwo}{{C}_2^n}
\newcommand{\rvckone}{{K}_1^n}
\newcommand{\rvcktwo}{{K}_2^n}
\newcommand{\rvcxone}{{X}^n}

\newcommand{\rvcxtwo}{{X}_2^n}
\newcommand{\rvcyone}{{Y}_1^n}
\newcommand{\rvcytwo}{{Y}_2^n}
\newcommand{\hrvcx}{\widehat{X}^n}
\newcommand{\hrvcxone}{\widehat{X}_1^n}
\newcommand{\hrvcxtwo}{\widehat{X}_2^n}

\newcommand{\lrvcc}{{C}^n}
\newcommand{\lrvck}{{K}^n}
\newcommand{\lrvcx}{{X}^n}
\newcommand{\lrvcy}{{Y}^n}
\newcommand{\lrvcz}{{Z}^n}
\newcommand{\lrvckone}{{K}_1^n}
\newcommand{\lrvcktwo}{{K}_2^n}

\newcommand{\lrvcxone}{{X}^n}
\newcommand{\lrvcxtwo}{{X}_2^n}
\newcommand{\lrvcyone}{{Y}_1^n}
\newcommand{\lrvcytwo}{{Y}_2^n}
\newcommand{\rvcci}{{C}_i^n}
\newcommand{\rvcki}{{K}_i^n}
\newcommand{\rvcxi}{{X}_i^n}
\newcommand{\rvcyi}{{Y}_i^n}
\newcommand{\hrvcxi}{\widehat{X}_i^n}
\newcommand{\vcki}{{k}_i^n}
\newcommand{\vcsi}{{s}_i^n}
\newcommand{\vcti}{{t}_i^n}
\newcommand{\vcvi}{{v}_i^n}
\newcommand{\vcwi}{{w}_i^n}
\newcommand{\vcxi}{{x}_i^n}
\newcommand{\vcyi}{{y}_i^n}

\newcommand{\vcs}{{s}^n}
\newcommand{\vct}{{t}^n}
\newcommand{\vcv}{{v}^n}
\newcommand{\vcw}{{w}^n}
}%%%%%%%%%%%%%%%%%%%%%%%%%%%%%%%%%%%%%%%%%%%%%%%%%%%%%%%

\newcommand{\crvcx}{\check{\vc X}}
\newcommand{\crvcxi}{\check{\vc X}_i}
\newcommand{\crvcxone}{\check{\vc X}_1}
\newcommand{\crvcxtwo}{\check{\vc X}_2}

   \newcommand{\lcrvcx}{\check{\vc X}}
\newcommand{\lcrvcxone}{\check{\lvc X}_1}
\newcommand{\lcrvcxtwo}{\check{\lvc X}_2}

  \newcommand{\cvcc}{\check{c}^m}
 \newcommand{\lcvcc}{\check{c}^m}
 \newcommand{\crvcc}{\check{C}^m}
\newcommand{\lcrvcc}{\check{C}^m}

  \newcommand{\cvccone}{\check{c}_1^{m_1}}
 \newcommand{\crvccone}{\check{C}_1^{m_1}}
  \newcommand{\cvcctwo}{\check{c}_2^{m_2}}
 \newcommand{\crvcctwo}{\check{C}_2^{m_2}}

\newcommand{\vca}{{\vc a}}
\newcommand{\vcb}{{\vc b}}
\newcommand{\vcc}{{\vc c}}
\newcommand{\vck}{{\vc k}}
\newcommand{\vcx}{{\vc x}}
\newcommand{\vcy}{{\vc y}}
\newcommand{\vcz}{{\vc z}}
\newcommand{\vckone}{{\vc k}_1}
\newcommand{\vcktwo}{{\vc k}_2}
\newcommand{\vcxone}{{\vc x}_1}
\newcommand{\vcxtwo}{{\vc x}_2}
\newcommand{\vcyone}{{\vc y}_1}
\newcommand{\vcytwo}{{\vc y}_2}

\newcommand{\cvcx}{\check{\vc x}}
\newcommand{\cvcy}{\check{\vc y}}
\newcommand{\cvcz}{\check{\vc z}}
\newcommand{\cvcxone}{\check{\vc x}_1}
\newcommand{\cvcxtwo}{\check{\vc x}_2}
\newcommand{\cvcxi}  {\check{\vc x}_i}

\newcommand{\hvcx}{\widehat{\vc x}}
\newcommand{\hvcy}{\widehat{\vc y}}
\newcommand{\hvcz}{\widehat{\vc z}}
\newcommand{\hvckone}{\widehat{\vc k}_1}
\newcommand{\hvcktwo}{\widehat{\vc k}_2}
\newcommand{\hvcki}  {\widehat{\vc k}_i}
\newcommand{\hvcxone}{\widehat{\vc x}_1}
\newcommand{\hvcxtwo}{\widehat{\vc x}_2}
\newcommand{\hvcxi}  {\widehat{\vc x}_i}

\newcommand{\lvca}{{\lvc a}}
\newcommand{\lvcb}{{\lvc b}}
\newcommand{\lvcc}{{\lvc c}}
\newcommand{\lvck}{{\lvc k}}
\newcommand{\lvcx}{{\lvc x}}
\newcommand{\lvcy}{{\lvc y}}
\newcommand{\lvcz}{{\lvc z}}

\newcommand{\lvckone}{{\lvc k}_1}
\newcommand{\lvcktwo}{{\lvc k}_2}

\newcommand{\lvcki}  {{\lvc k}_i}
\newcommand{\lvcxi}  {{\lvc x}_i}
\newcommand{\lvcxone}{{\lvc x}_1}
\newcommand{\lvcxtwo}{{\lvc x}_2}

\newcommand{\lvcyone}{{y}_1}
\newcommand{\lvcytwo}{{y}_2}

\newcommand{\clvcxone}{\check{\lvc x}_1}
\newcommand{\clvcxtwo}{\check{\lvc x}_2}
\newcommand{\clvcxi}{\check{\lvc x}_i}

\newcommand{\hlvckone}{\widehat{k}_1}
\newcommand{\hlvcktwo}{\widehat{k}_2}
\newcommand{\hlvcxone}{\widehat{x}_1}
\newcommand{\hlvcxtwo}{\widehat{x}_2}

\newcommand{\rvcc}{{\vc C}}
\newcommand{\rvck}{{\vc K}}
\newcommand{\rvcx}{{\vc X}}

\newcommand{\rvcy}{{\vc Y}}
\newcommand{\rvcz}{{\vc Z}}
\newcommand{\rvccone}{{\vc C}_1}
\newcommand{\rvcctwo}{{\vc C}_2}
\newcommand{\rvckone}{{\vc K}_1}
\newcommand{\rvcktwo}{{\vc K}_2}

\newcommand{\rvcxone}{{\vc X}_1}
\newcommand{\rvcxtwo}{{\vc X}_2}
\newcommand{\rvcyone}{{\vc Y}_1}
\newcommand{\rvcytwo}{{\vc Y}_2}
\newcommand{\hrvcxone}{\widehat{\vc X}_1}
\newcommand{\hrvcxtwo}{\widehat{\vc X}_2}

\newcommand{\lrvcc}{{\lvc C}}
\newcommand{\lrvck}{{\lvc K}}
\newcommand{\lrvcx}{{\lvc X}}
\newcommand{\lrvcy}{{\lvc Y}}
\newcommand{\lrvcz}{{\lvc Z}}

\newcommand{\lrvcxi  }{{\lvc X}_i}
\newcommand{\lrvcki  }{{\lvc K}_i}
\newcommand{\lrvckone}{{\lvc K}_1}
\newcommand{\lrvcktwo}{{\lvc K}_2}
\newcommand{\lrvcxone}{{\lvc X}_1}
\newcommand{\lrvcxtwo}{{\lvc X}_2}
\newcommand{\lrvcyone}{{\lvc Y}_1}
\newcommand{\lrvcytwo}{{\lvc Y}_2}

\newcommand{\srvcc}{{\svc C}}
\newcommand{\srvck}{{\svc K}}
\newcommand{\srvcx}{{\svc X}}
\newcommand{\srvcy}{{\svc Y}}
\newcommand{\srvcz}{{\svc Z}}

\newcommand{\srvcxi  }{{\svc X}_i}
\newcommand{\srvckone}{{\svc K}_1}
\newcommand{\srvcktwo}{{\svc K}_2}
\newcommand{\srvcxone}{{\svc X}_1}
\newcommand{\srvcxtwo}{{\svc X}_2}
\newcommand{\srvcyone}{{\svc Y}_1}
\newcommand{\srvcytwo}{{\svc Y}_2}

\newcommand{\rvcci}{{\vc C}_i}
\newcommand{\rvcki}{{\vc K}_i}
\newcommand{\rvcxi}{{\vc X}_i}
\newcommand{\rvcyi}{{\vc Y}_i}
\newcommand{\hrvcxi}{\widehat{\vc X}_i}
\newcommand{\vcki}{{\vc k}_i}
\newcommand{\vcsi}{{\vc s}_i}
\newcommand{\vcti}{{\vc t}_i}
\newcommand{\vcvi}{{\vc v}_i}
\newcommand{\vcwi}{{\vc w}_i}
\newcommand{\vcxi}{{\vc x}_i}
\newcommand{\vcyi}{{\vc y}_i}

\newcommand{\vcs}{{\vc s}}
\newcommand{\vct}{{\vc t}}
\newcommand{\vcv}{{\vc v}}
\newcommand{\vcw}{{\vc w}}
%\newcommand{\vcx}{{\vc x}}
%\newcommand{\vcy}{{\vc y}}
%%%%%%%%%%%%%%%%%%%%%%%%%%%%%%%%%%%%%%%%%%%%%%%%%%%%%%%
%
%

\newcommand{\CommenT}{%%%%%%%%%%%%%%%%%%%%%%%%%%%%%%%%%%%
%%%%%%%%%%%%%%%%%%%%%%%%%%%%%%%%%%%%%%%%%%%%%%%%%%%%%%%%% 

First, allow us to restate the main results of our paper.

We propose a new security metric for encryption schemes which is more
strict compared with the widely-used security metric based on the
mutual information. Let $\Delta$ denote the newly proposed metric and
$I$ denote the mutual information between the plaintexts and the 
ciphertexts. The new metric denoted by $\Delta$ is valid in the sense
that if $I=0$, then $\Delta=0$. This result stated in Property part a)
is the most essential part of our results. We prove the necessary and
sufficient condition to achieve the security for distributed
encryption with correlated keys under the newly proposed metric.

The extended version of this paper with a more detailed explanation is
presented in the following arXiv paper.
Yasutada Oohama, Bagus Santoso,
``Distributed Source Coding with Encryption Using Correlated Keys",
arXiv:2102.06363 [cs.IT].

Below, we will explain the significance of our results in brief.

In the general case, proving the necessary condition under own
proposed metric is more strict than the widely-used metric is
meaningless. However, we claim that our newly proposed metric is a
rare example of a special case that does not fall into the trap of the
general case above.

The first reason is that we already have a concrete scheme that
satisfies the more strict requirement of the proposed metric, i.e.,
the distributed encryption scheme proposed by Santoso and Oohama [1],
[2]. In [1],[2], on the surface, Santoso and Oohama stated that their
distributed encryption scheme is secure under the security metric
based on the mutual information. However, if one looks underneath a
little bit, one can easily discover that actually Santoso and Oohama
proved the security of their scheme based on a metric that is more
strict than the mutual information. And one can easily see that the
more strict security metric they used is equal to our new proposed
metric. Thus, our proposed metric is not a mere theoretical concept,
but an achievable security requirement in the practical world.

The second reason is that we argue that the proposed metric $\Delta$
has a deeper relationship with the already widely-used security metric
in information theory, i.e., the mutual information $I$.  In this
paper, based on our new proposed extension of the Birkhoff-von Neumann
theorem (Lemma 1), we have proven that for any decodable encryption
scheme, the following holds: $I=0$ only if $\Delta=0$ (Property 3(a)).
In other words, an encryption scheme can not be proven to be perfectly
secure based on the mutual information ($I=0$) unless it is perfectly
secure under our new proposed metric ($\Delta=0$). Although we have
not shown the general case where $\Delta$ and $I$ is nearly zero in
this paper yet, since we have shown that the newly proposed metric is
not only acting as the theoretical "upper-bound" of the mutual
information but is also acting as the "lower-bound" of it in the
perfect case, we argue that one can treat our proposed metric as a
valid alternative of security metric in information theory.

Here we will explain another meaning of the statement "for any
decodable encryption: $I=0$ only if $\Delta=0$" (Property 3(a)). As we
state in the paper, one can represent the new proposed metric $\Delta$
as a direct sum of (1) the mutual information $I$ between plaintexts
and ciphertexts, and (2)the divergence between the distribution of
ciphertexts and uniform distribution. Informally speaking, one can see
$\Delta$ as the summation of the degree of dependency between
plaintexts and ciphertexts and the distance between the ciphertexts'
distribution and uniform distribution. Hence, for any encryption
scheme, unless the distribution of the ciphertexts is completely
uniform, the plaintexts and ciphertexts can not be completely
independent. In other words, since the complete independence between
plaintexts and ciphertexts means perfect security, one can also say
that unless one flattens the distribution of ciphertexts into a
completely uniform distribution, one can not achieve perfect security.

Now we will explain the relationship between the above statement of
Property 3(a) and the result of Santoso and Oohama in [1],[2]. In [1],
[2], Santoso and Oohama used the mutual information as the security
metric. Informally, Santoso and Oohama use a certain compression
function to flat the distribution of ciphertexts in each encryption
node before they are released to the public channel such that even in
the case that the encryption keys between distributed sources are
correlated, the security is guaranteed. However, it has not been clear
whether the flattening of the ciphertexts is the only possible method
to guarantee the security of the scheme. In short, the above statement
of Property 3(a) said that in the perfect case, the flattening of the
ciphertexts into a perfect uniform distribution is the only way to
make the scheme achieve perfect security.
%%%%%%%%%%%%%%%%%%%%%%%%%%%%%%%%%%%%%%%%%%%%%%%%%%%%%%%%%%
}%%%%%%%%%%%%%%%%%%%%%%%%%%%%%%%%%%%%%%%%%%%%%%%%%%%%%%%%%
%%%%%%%%%%%%%%%%%%%%%%%%%%%%%%%%%%%%%%%%%%%%%%%%%%%%%%%%%%
%%%%%%%%%%%%%%%%%%%%%%%%%%%%%%%%%%%%%%%%%%%%%%%%%%%%%%%%%%

\section{Introduction \label{sec:introduction}}

In this paper we pose and investigate the distributed secure source 
coding based on the common key cryptosystem. This cryptosystem 
includes the secrecy amplification problem for distributed encrypted 
sources with correlated keys using post-encryption-compression 
(PEC), which was posed investigated by Santoso and Oohama 
in \cite{DBLP:conf/isit/SantosoO17}, \cite{santosoOhPEC:19}. 

%We have 
%two main results in our paper. One is that 

In this paper we propose a new security criterion 
which is more natural compared with the commonly 
used security criterion based on the upper-bound
of mutual information between the plaintext and the ciphertext. 
For the proposed new metric we prove its {\it validity}. 
Concretely we establish the following two results:
\begin{itemize}
\item[a)] 
We prove that if the mutual information is zero, then 
the proposed criterion is strictly zero. 
\item[b)]
The proposed criterion depends only on the property of 
the cryptsystem, implying that 
this criterion is more natural than the 
widely-used security metric based mutual information. 
\end{itemize}
For the proposed security criterion, the part a) is quite essential. 
Without this condition, the criterion is meaningless. 

%Under this criterion, we establish the necessary and sufficient 
%condition for the secure transmission of correlated sources.
%In this paper we propose a new valid 
%security metric for encryption schemes which is more strict compared 
%with the widely-used security metric based on the mutual information. 
%Under the proposed security criterion 

Under the proposed security criterion we prove the strong converse 
theorem. 
%This result is stated in Theorem \ref{th:SStConvThTwo}.
%Due to the introduction of this criterion the proof of the strong 
%converse theorem is quite simple. 
We further derive a sufficient 
condition to achieve security. This 
sufficient condition matches the necessary condition. 
In our previous works of 
Santoso and Oohama \cite{DBLP:conf/isit/SantosoO17}, 
                   \cite{santosoOhPEC:19}, we have 
derived a sufficient condition under the security criterion 
measured by the mutual information. To derive the sufficient 
condition we use the coding scheme proposed by Santoso and Oohama. 
We obtain the same sufficient condition 
as that of Santoso 
and Oohama \cite{DBLP:conf/isit/SantosoO17}, 
           \cite{santosoOhPEC:19}
under more natural condition than the mutual information. 

Our study in this paper has a closely related to 
several previous works on the PEC, e.g., Johnson et al.\cite{1337277},
Klinc et al. \cite{DBLP:journals/tit/KlincHJKR12}.
Our study also has a close connection with several previous works 
on the Shannon cipher system, e.g. \cite{Sh48}, \cite{Ya91} 
\cite{DBLP:journals/tit/IwamotoOS18}.

\section{Secure Source Coding Problem}

\subsection{Preliminaries}

In this subsection, we show the basic notations and related consensus 
used in this paper. 

\ \par \noindent{}\textit{Random Sources of Information and Keys: \ }
Let $(X_1,X_2)$ be a pair of random variables from a finite set 
$\mathcal{X}_1\times \mathcal{X}_2$. 
%and $(K_1,K_2)$ be a pair of random variables from a finite set 
%$\mathcal{K}_1,\times\mathcal{K}_2$.
Let $\{(X_{1,t},X_{2,t})\}_{t=1}^\infty$ be a stationary \emph{discrete
memoryless source} (DMS) such that for each $t=1,2,\ldots$, 
the pair $(X_{1,t},X_{2,t})$ takes values in finite set 
$\mathcal{X}_1\times \mathcal{X}_2$ and obeys the same distribution 
as that of $(X_1,X_2)$ denoted by 
$p_{X_1 X_2}=\{p_{X_1 X_2} (x_1,x_2)\}_{(x_1,x_2)
\in\mathcal{X}_1\times 
\mathcal{X}_2}$. The stationary DMS 
$\{(X_{1,t},X_{2,t})\}_{t=1}^\infty$ 
is specified with $p_{X_1X_2}$. %Similar notations are adopted for 
%other random variables.
\newcommand{\Zap}{}{%%%%%%%%%%%%%%%%%%%%%%%%%%%%%%
Also, let $(K_1, K_2)$ be a pair of random variables taken from 
the same  finite set $\mathcal{X}_1\times \mathcal{X}_2$ 
representing the pair of keys used for encryption at 
two separate terminals, of which the detailed description will 
be presented later. Similarly, 
let $\{(K_{1,t},K_{2,t})\}_{t=1}^\infty$ 
be a stationary discrete
memoryless source such that for each $t=1,2,\ldots$, 
the pair $(K_{1,t},K_{2,t})$ takes values in finite set
$\mathcal{X}_1\times \mathcal{X}_2$ and obeys the same 
distribution as that of $(K_1,K_2)$ denoted by 
$p_{K_1 K_2}=
\{p_{K_1 K_2} (k_1,k_2)\}_{(k_1,k_2)
\in\mathcal{X}_1\times \mathcal{X}_2}$.
The stationary DMS $\{(K_{1,t},K_{2,t})\}_{t=1}^\infty$ 
is specified with $p_{K_1K_2}$.
}%%%%%%%%%%%%%%%%%%%%%%%%%%%%%%%%%%%%%%%%%%%%%%%%%

\ \par\noindent{}\textit{Random Variables and Sequences: \ }
We write the sequence of random variables with length $n$ 
from the information source as follows:
${\rvcxone}\defeq X_{1,1}X_{1,2}\cdots X_{1,n}$, 
${\rvcxtwo}\defeq X_{2,1}X_{2,2}\cdots X_{2,n}$.
Similarly, the strings with length $n$ of $\mathcal{X}_1^n$ 
and $\mathcal{X}_2^n$ are written as 
${\vcxone}\defeq x_{1,1}x_{1,2}\cdots x_{1,n}\in\mathcal{X}_1^n$ 
and ${\vcxtwo}
\defeq
x_{2,1}x_{2,2}\cdots x_{2,n}\in\mathcal{X}_2^n$ respectively.
For $({\vcxone}, {\vcxtwo})\in\mathcal{X}_1^n\times \mathcal{X}_2^n$, 
$p_{{\lrvcxone}{\lrvcxtwo}}({\vcxone},{\vcxtwo})$ stands for the 
probability of the occurrence of $({\vcxone}, {\vcxtwo})$. 
When the information source is memoryless 
specified with $p_{X_1X_2}$, we have the following equation holds:
$p_{{\lrvcxone} {\lrvcxtwo}}({\vcxone},{\vcxtwo})=
\prod_{t=1}^n p_{X_1X_2}(x_{1,t},x_{2,t})$.
In this case we write $p_{{\lrvcxone} {\lrvcxtwo}}({\vcxone},{\vcxtwo})$
as $p_{X_1 X_2}^n({\vcxone},{\vcxtwo})$.
Similar notations are used for other random variables and sequences.

\ \par\noindent{}\emph{Consensus and Notations: }
Without loss of generality, throughout this paper,
we assume that $\mathcal{X}_1$ and $\mathcal{X}_2$ are finite fields.
The notation $\oplus$ is used to denote the field addition operation,
while the notation $\ominus$ is used to denote the field subtraction 
operation, i.e., $a\ominus b = a \oplus (-b)$ for any
elements $a,b$ of a same finite field. 
%All discussions and 
%theorems in this paper still hold although
%$\mathcal{X}_1$ and $\mathcal{X}_2$ are different finite
%fields. However, 
For the sake of simplicity, we use the same notation for field 
addition and subtraction for both $\mathcal{X}_1$ and $\mathcal{X}_2$.
Throughout this paper all logarithms are taken to the base 2. 

\subsection{Basic System Description}

First, let the information sources and keys be generated 
independently by different parties 
$\Sgen$ and $\Kgen$ respectively.
In our setting, we assume the followings.
\begin{itemize}
	\item The random keys ${\rvckone}$  and ${\rvcktwo}$ 
              are generated by $\Kgen$.
	      %from uniform distribution.
	\item The key ${\rvckone}$ is correlated to  
	      ${\rvcktwo}$. 
	\item The sources ${\rvcxone}$ and ${\rvcxtwo}$ are 
              generated by $\Sgen$ and are correlated
	      to each other.
	\item The sources are independent to the keys.
\end{itemize}

\begin{figure}[t]
\centering
\includegraphics[width=0.47 \textwidth]{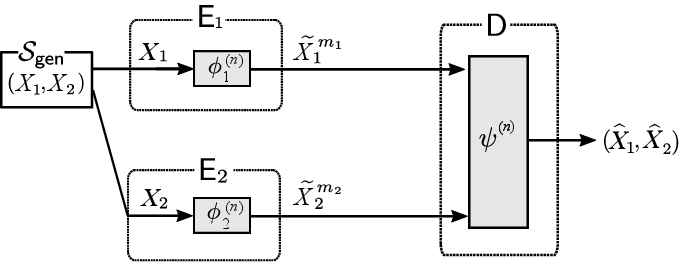}
\caption{Distributed source coding without encryption.
\label{fig:mainA}}
\end{figure}

\noindent
\underline{\it Source coding without encryption:} \ The 
two correlated random sources ${\rvcxone}$ and ${\rvcxtwo}$ 
from $\Sgen$ be sent to two separated nodes $\mathsf{E}_1$ 
and $\mathsf{E}_2$ respectively.
Further settings of the system are described as follows. 
Those are also shown in Fig. \ref{fig:mainA}.

\begin{enumerate}
	\item \emph{Encoding Process:} \ 
        For each $i=1,2$, at the node $\mathsf{E}_i$, 
        the encoder function 
        $\phi_{i}^{(n)}: {\cal X}_i^n $ $\to {\cal X}_i^{m_i}$ 
        observes ${\rvcxi}$ to generate 
        $\tilde{X}_i^{m_i}=\phi_{i}^{(n)}({\rvcxi})$. 
        Without loss of generality we may assume that 
        $\phi_{i}^{(n)}$ is {\it surjective}. 
	\item \emph{Transmission:} \ 
        Next, the encoded sources $\tilde{X}_i^{m_i}$, $i=1,2$ 
        are sent to the 
        information processing center $\D$ through two \emph{noiseless} 
        channels. 
	\item \emph{Decoding Process:} \ 
        In $\D$, the decoder function observes $\tilde{X}^{m_i},i=1,2$ 
        to output $(\widehat{\rvcx}_1,\widehat{\rvcx}_2)$,
        using the one-to-one mapping $\psi^{(n)}$ defined by 
        $\psi^{(n)}:
             {\cal X}_1^{m_1}\times {\cal X}_2^{m_2} 
         \to {\cal X}_1^n    \times {\cal X}_2^n$. 
         Here we set 
        \begin{align*}
         (\widehat{\rvcx}_1,\widehat{\rvcx}_2) \defeq & 
        \psi^{(n)}(\tilde{X}^{m_1}_1,
                             \tilde{X}^{m_2}_2)
         \\ 
         =& \psi^{(n)}\left(
         \phi^{(n)}_{1}(\rvcxone),
         \phi^{(n)}_{2}(\rvcxtwo)\right).
        \end{align*}
        %Without loss of generality we assume 
        %that $\psi_{\bullet}^{(n)}$
        %is a {\it one-to-one mapping}. 
More concretely, the decoder outputs the unique pair 
$(\widehat{\rvcx}_1,\widehat{\rvcx}_2)$ from 
$(\phi_{1}^{(n)})^{-1}(\tilde{X}_1^{m_1}) \times 
 (\phi_{2}^{(n)})^{-1}(\tilde{X}_2^{m_2})$ 
in a proper manner.
\end{enumerate}
%%%%%%%%%%%%%%%%%%%%%%%%%%%%%%%%%%%%%%%%%%%%%%%%%%%%%%
%%%%%%%%%%%%%%%%%%%%%%%%%%%%%%%%%%%%%%%%%%%%%%%%%%%%%%

\newcommand{\Omittbbb}{%%%%%%%%%%%%
}{%%%%%%%%%%%%%%%%%%%%%%%%%%%%%%%%%
%%%%%%%%%%%%%%%%%%%%%%%%%%%%%%%%%%%
\begin{figure}[t]
\centering
\includegraphics[width=0.47 \textwidth]{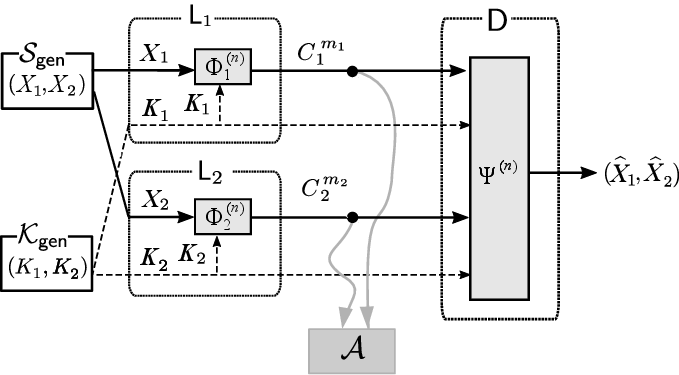}
\caption{Distributed source coding with encryption.
\label{fig:main}}
\end{figure}
}%%%%%%%%%%%%%%%%%%%%%%%%%%%%%%%%%
%%%%%%%%%%%%%%%%%%%%%%%%%%%%%%%%%%

%Since the mapping $\psi_{0}^{(n)}:\mathcal{X}^{m}\to\mathcal{X}^{n}$
%is defined as a one-to-one mapping, it is easy to see that the above
%source coding system satisfies the following lemma.
%\begin{lem}%\vspace{-3mm}
%Let 
For the above $(\phi_1^{(n)},\phi_2^{(n)},\psi^{(n)})$, we define 
the set $ \mathcal{D}^{(n)}$ of correct decoding by 
\begin{align*}
\mathcal{D}^{(n)} & :=
\{(\vcxone,\vcxtwo) \in \mathcal{X}_1^{n}\times \mathcal{X}_2^{n}:
\\ 
&\psi^{(n)}( 
\varphi_{1}^{(n)}(\vcxone),
\varphi_{2}^{(n)}(\vcxtwo))
=(\vcxone, \vcxtwo)\}.
\end{align*}
On $|{\cal D}^{(n)}|$, we have the following property. 
\begin{property}\label{pr:prOnDecSet} 
We have the following.  
\begin{align}
& |{\cal D}^{(n)}|=|{\cal X}_1^{m_1}||{\cal X}_2^{m_2}|. 
\label{eqn:CardOne}
\end{align}
\end{property}

Proof of Property \ref{pr:prOnDecSet} is given 
in Appendix \ref{apd:ProofPrOnDecSet}. 
\begin{remark}
In brief, the reason that we can assume the decoder as injective
mapping without loss of generality is that for any non-injective
decoder, we can construct an injective decoder with the same
performance. More concretely, for any encoder
$\tilde{\phi}^{(n)}:\mathcal{X}_i^n 
\rightarrow \mathcal{X}_i^{m_i}, i=1,2$ 
and any $\tilde{\psi}^{(n)}$ 
not necessary injective, there exists
$(\phi_1^{(n)},\phi_2^{(n)},\psi^{(n)})$ where 
$\psi^{(n)}$ is injective such that the following holds:
$\tilde{\psi}^{(n)}(\tilde{\phi}_1({\cal X}_1^n),
 \tilde{\phi}_2({\cal X}_2^n))=\psi^{(n)}(\phi_1({\cal X}_1^n),
\phi_2({\cal X}_2^n))$, 
$|\tilde{\phi}_i({\cal X}_i^n)|\geq 
 |      {\phi}_i({\cal X}_i^n)|,i=1,2$.
\end{remark}

\newcommand{\ProofPrOnDecSet}{%%%%%%%%%%%%%%%%%%
\subsection{
%Properties on the Decoding Sets 
Proof of Property \ref{pr:prOnDecSet} 
}
\label{apd:ProofPrOnDecSet}
%In this appendix we prove the property on 
%the decoding set ${\cal D}^{(n)}$ stated 
%in Property \ref{pr:prOnDecSet}. 
\begin{IEEEproof}[Proof of Property \ref{pr:prOnDecSet}] 
We have the following:
\begin{align}
{\cal D}^{(n)}\MEq{a}&
\{(\vcxone,\vcxtwo)
=\psi^{(n)}(\tilde{x}_1^{m_1},\tilde{x}_2^{m_2}):
\notag\\
& \quad (\tilde{x}_1^{m_1}, \tilde{x}_2^{m_2}) 
\in \phi_1^{(n)}({\cal X}_1^{n}) \times 
    \phi_2^{(n)}({\cal X}_2^{n}) \}
\notag\\
\MEq{b}&
\{(\vcxone,\vcxtwo)=\psi^{(n)}
(\tilde{x}_1^{m_1},
 \tilde{x}_2^{m_2}):
\notag\\
& \quad (\tilde{x}_1^{m_1}, 
         \tilde{x}_2^{m_2}) 
         \in {\cal X}_1^{m_1} \times {\cal X}_2^{m_2} \}.
\label{eqn:pSddxcc} 
\end{align}
Step (a) follows from that every pair 
$
(\tilde{x}_1^{m_1}, \tilde{x}_2^{m_2})  \in 
\phi_1^{(n)}({\cal X}_1^{n}) \times 
\phi_2^{(n)}({\cal X}_2^{n}) \}$ uniquely 
determines $(\vcxone,\vcxtwo)\in {\cal D}^{(n)}$.
Step (b) follows from that $\phi_i^{(n)},i=1,2$ are surjective.
Since 
$\psi^{(n)}: {\cal X}_1^{m_1} \times {\cal X}_2^{m_2}
\to {\cal X}_1^n \times {\cal X}_2^n$ is a one-to-one mapping 
and (\ref{eqn:pSddxcc}), 
we have 
$
|{\cal D}^{(n)}|=|{\cal X}_1^{m_1}||{\cal X}_2^{m_2}|.
$
\end{IEEEproof}
}%%%%%%%%%%%%%%%%%%%%%%%%%%%%%%%%%%%%%%%%%%%%%%%%%%%%%
%\in {\cal X}_1^{m_1}\times \in {\cal X}_2^{m_2}
%%%%%%%%%%%%%%%%%%%%%%%%%%%%%%%%%%%%%%%%%%%%%%%%%%%%%%
%%%%%%%%%%%%%%%%%%%%%%%%%%%%%%%%%%%%%%%%%%%%%%%%%%%%%%

%\end{document}

%%%%%%%%%%%%%%%%%%%%%%%%%%%%%%%%%%%%%%%%%%%%%%%%%%%%%%%%%%%%%%%%%%%%%%%%%%
%Then, the following holds: $|\mathcal{D}_{n}|=|\mathcal{X}^{m}|$.
%\end{lem}
% {\it Condition:}  
%The following property basically says that the input and the result
%of the whole process of encryption and decryption of the cryptosystem
%which involves the secret key and the plaintext can be rewritten as
%a simple process of source coding system in the main building block
%which involves only the plaintext.
%%%%%%%%%%%%%%%%%%%%%%%%%%%%%%%%%%%%%%%%%%%%%%%%%%%%%%%%%%%%%%%%%%%%%%%%%%
%%%%%%%%%%%%%%%%%%%%%%%%%%%%%%%%%%%%%%%%%%%%%%%%%%%%%%%%%%%%

\noindent
\underline{\it Distributed source coding with encryption:} \ 

The two correlated random sources ${\rvcxone}$ and ${\rvcxtwo}$ 
from $\Sgen$ are sent to two separated nodes 
$\mathsf{L}_1$ and $\mathsf{L}_2$, respectively. 
The two random keys ${\rvckone}$ and ${\rvcktwo}$ from $\Kgen$, 
are also sent to $\mathsf{L}_1$ and and $\mathsf{L}_2$, respectively. 
Further settings of our system are described as follows. 
%%%%%%%%%%%%%%%%%%%%%%%%%%%%%%%%%%%%%%%%%%%%%%%
Those are also shown in Fig. \ref{fig:main}.
%%%%%%%%%%%%%%%%%%%%%%%%%%%%%%%%%%%%%%%%%%%%%%
\begin{enumerate}
\item \emph{Source Processing:} \ For each $i=1,2$, 
at the node $\L_i$, ${\rvcx}_i$ is encrypted with the key ${\rvck}_i$ 
using the encryption function 
        $\Phi_i^{(n)}:{\cal X}_i^n \times {\cal X}_i^n$ 
       	$\to {\cal X}_i^{m_i}$. 
        For each $i=1,2$, the ciphertext $C_i^{m_i}$ of ${\rvcx}_i$ 
        is given by $C_i^{m_i}=\Phi_{i}^{(n)}
         ({\rvck}_i,{\rvcx}_i)$. 
        On the encryption function $\Phi_i^{(n)}, i=1,2$, we 
        use the folloiwng notation:
        $$
         \Phi_i^{(n)}({\rvck}_i, {\rvcx}_i)
        =\Phi^{(n)}_{i, {\lrvck}_i}({\rvcx}_i)
        =\Phi^{(n)}_{i, {\lrvcx}_i}({\rvck}_i).
        $$  
        %%%%%%%%%%%%%%%%%%%%%%%%%%%%%%%%%%%%%%%%%%%%%%%%%%%%%%%%%%%%
        \item \emph{Transmission:} \ Next, the ciphertext 
        $C_i^{m_i},i=1,2$ 
        are sent to the information processing center $\D$ through 
        two \emph{public} communication channels. 
        Meanwhile, the key ${\rvck}_i, i=1,2$, are sent 
        to $\D$ through two \emph{private} communication channels.
	%%%%%%%%%%%%%%%%%%%%%%%%%%%%%%%%%%%%%%%%%%%%%%%%%%%%%%%%%%%%
        \item \emph{Sink Node Processing:} \ In $\D$, we decrypt 
        the ciphertext $(\widehat{\rvcx}_1,\widehat{\rvcx}_2)$ 
         from $C_i^{m_i},i=1,2,$ using 
        the key ${\rvck}_i,i=1,2$, through the corresponding 
         decryption procedure $\Psi^{(n)}$ defined by 
        $ 
         \Psi^{(n)}:
         {\cal X}_1^n \times {\cal X}_2^n \times 
         {\cal X}_{1}^{m_1} \times {\cal X}_{2}^{m_2} 
         \to {\cal X}_1^n \times {\cal X}_2^n.
        $
        Here we set 
        $$
         (\widehat{\rvcx}_1,\widehat{\rvcx}_2)\defeq 
          \Psi^{(n)}({\rvckone},{\rvcktwo},C_1^{m_1},C_2^{m_2}).
        $$ 
        More concretely, the decoder outputs the unique pair 
        $(\widehat{\rvcx}_1,\widehat{\rvcx}_2)$ from 
        $(\Phi_{1,\lrvckone}^{(n)})^{-1}({C}_1^{m_1}) \times 
         (\Phi_{2,\lrvckone}^{(n)})^{-1}({C}_2^{m_2})$ 
        in a proper manner. On the decryption function 
        $\Psi^{(n)}$, we use the following notation:
        \begin{align*}
        & \Psi^{(n)}({\rvck}_1,{\rvck}_2,C_1^{m_1},C_2^{m_2})
        =\Psi^{(n)}_{\lrvckone,\lrvcktwo}(C_1^{m_1},C_2^{m_2})
        \\
        &=\Psi^{(n)}_{C_1^{m_1},C_2^{m_2}}
          ({\rvckone},{\rvcktwo}).
        \end{align*}  
\end{enumerate}

Fix any $(\rvckone, \rvcktwo)=(\vckone,\vcktwo) 
\in \mathcal{X}_1^{n}\times \mathcal{X}_2^{n}$.
For this $(\rvckone,\rvcktwo)$ and 
for $(\Phi_1^{(n)},\Phi_2^{(n)},\Psi^{(n)})$,  
we define the set $ \mathcal{D}^{(n)}_{\lvckone,\lvcktwo}$ 
of correct decoding by 
\begin{align*}
\mathcal{D}^{(n)}_{\lvckone,\lvcktwo}
& := \{(\vcxone,\vcxtwo) \in \mathcal{X}_1^{n}
                      \times \mathcal{X}_2^{n}:
\\ 
&\Psi^{(n)}%\circ 
(\Phi_{1}^{(n)}(\vckone, \vcxone),
(\Phi_{2}^{(n)}(\vcktwo, \vcxtwo))
=(\vcxone, \vcxtwo)\}.
\end{align*}
We require that the cryptosystem 
$
(\Phi_{1}^{(n)}, 
 \Phi_{2}^{(n)},
 \Psi^{(n)})
$
must satisfy the following condition.

{\it Condition:} 
For each distributed source encryption sysytem 
$(\Phi_{1}^{(n)},\Phi_{2}^{(n)},\Psi^{(n)})$,  
there exists a distributed source coding system %specified with 
$(\phi_1^{(n)}, \phi_2^{(n)}, \psi^{(n)})$ such that  
for any $(\vckone, \vcktwo) \in \mathcal{X}_1^{n} 
\times \mathcal{X}_2^{n}$ and for any $(\vcxone, \vcxtwo) 
\in \mathcal{X}_1^{n} \times \mathcal{X}_2^{n}$, 
%the following holds:
\begin{align*}
& \Psi_{\lvckone,\lvcktwo}^{(n)}%\circ 
 (\Phi_{1,\lvckone}^{(n)}(\vcxone),
  \Phi_{2,\lvcktwo}^{(n)}(\vcxtwo)) 
\\
& =\psi^{(n)}%\circ
(\phi_{1}^{(n)}(\vcxone), 
                   \phi_{2}^{(n)}(\vcxtwo)). 
\end{align*}
The above condition implies that 
$$
{\cal D}^{(n)}={\cal D}^{(n)}_{\lvckone,\lvcktwo}, 
\forall (\vckone,\vcktwo) \in {\cal X}_1^n\times {\cal X}_2^n.
$$

We have the following properties on ${\cal D}^{(n)}$. 
\begin{property}
\label{pr:prOne}
\begin{itemize}
\item[a)] 
If $(\vcxone,\vcxtwo), 
    (\vcx^{\prime}_1,
     \vcx^{\prime}_2) 
\in {\cal D}^{(n)}$ and $(\vcxone,\vcxtwo) 
\neq (\vcx^{\prime}_1,\vcx^{\prime}_2)$, then 
$$
(\Phi^{(n)}_{1,\lvckone}(\vcxone), 
 \Phi^{(n)}_{2,\lvcktwo}(\vcxtwo))
\neq 
(\Phi^{(n)}_{1,\lvckone}(\vcx^{\prime}_1), 
 \Phi^{(n)}_{2,\lvcktwo}(\vcx^{\prime}_2)).
$$
\item[b)] %
    $\forall (\vckone,\vcktwo)$ 
and $\forall (c_1^{m_1},c_2^{m_2})$, 
$\exists (\vcxone,\vcxtwo)$ $\in {\cal D}^{(n)}$ such that  
$$
(\Phi^{(n)}_{1,\lvckone}(\vcxone), 
 \Phi^{(n)}_{2,\lvcktwo}(\vcxtwo))
=(c_1^{m_1}, 
  c_2^{m_2}).
$$
\end{itemize}
\end{property}

Proof of Property \ref{pr:prOne} is given 
in Appendix \ref{apd:ProofPrOne}. 
\newcommand{\ProofPrOne}{%%%%%%%%%%%%%%%%%%%%%%%%%%%%%%%%%%%%%%
\subsection{
Proof of Property \ref{pr:prOne}
}\label{apd:ProofPrOne}
%}{

%In this appendix we prove Property \ref{pr:prOne}.
We frist prove the part a) and next prove the part b).   
%%%%%%%%%%%%%%%%%%%%%%%%%%%%%%%%%%%%%%%%%%%%%%%%%%%%%%%%%%%%%%
%%%%%%%%%%%%%%%%%%%%%%%%%%%%%%%%%%%%%%%%%%%%%%%%%%%%%%%%%%%%%%
\begin{IEEEproof}[Proof of Property \ref{pr:prOne} part a)] 
Under $(\vcxone,\vcxtwo), (\vcx_1^{\prime},\vcx_2^{\prime}) 
\in {\cal D}^{(n)}$ and 
$(\vcxone,\vcxtwo)\neq$ 
$(\vcx_1^{\prime},\vcx_2^{\prime})$, 
we assume that 
\beq
 (\Phi^{(n)}_{1,\lvckone}({\vcxone}), 
  \Phi^{(n)}_{2,\lvcktwo}({\vcxtwo})) 
=(\Phi^{(n)}_{1,\lvckone}({\vcx}_1^{\prime}), 
  \Phi^{(n)}_{2,\lvcktwo}({\vcx}_2^{\prime})).
\label{eqn:Assum} 
\eeq
Then  we have the following 
\begin{align}
&(\vcxone,\vcxtwo)\MEq{a}
\psi^{(n)}
(\phi_1^{(n)}(\vckone),
           \phi_2^{(n)}(\vcktwo) ,
\notag\\
&\MEq{b}
\Psi^{(n)}_{\lvckone,\lvcktwo}
(\Phi^{(n)}_{1,\lvckone}({\vcxone}),  
 \Phi^{(n)}_{2,\lvcktwo}({\vcxtwo}))
\notag\\
& \MEq{c}
\Psi^{(n)}_{\lvckone,\lvcktwo}
(\Phi^{(n)}_{1,\lvckone}({\vcx}_1^{\prime}),
 \Phi^{(n)}_{2,\lvcktwo}({\vcx}_2^{\prime}))
\notag\\
& \MEq{d}
 \psi^{(n)}
(\phi^{(n)}_1({\vcx}_1^{\prime}),
 \phi^{(n)}_2({\vcx}_2^{\prime}))\MEq{e}
             ({\vcx}_1^{\prime},
              {\vcx}_2^{\prime}).
\label{eqn:SdCCv}
\end{align}
Steps (a) and (e) follow from the definition of 
${\cal D}^{(n)}$. Step (c) follows from \ref{eqn:Assum}.
Steps (b) and (d) follow from the relationship between 
$
(\phi^{(n)}_1,
 \phi^{(n)}_2,
 \psi^{(n)})
$
and 
$
(\Phi^{(n)}_{1,\lvckone},
 \Phi^{(n)}_{2,\lvcktwo},
 \Psi^{(n)}_{  \lvckone,
               \lvcktwo}).$
The equality (\ref{eqn:SdCCv}) contradics the first assumption.
Hence we must have Property \ref{pr:prOne} part a).
\end{IEEEproof}
\begin{IEEEproof}[Proof of Property \ref{pr:prOne} part b)] 
We assume that $ \exists (\vckone,\vcktwo)$ and 
$\exists (c_1^{m_1},c_2^{m_2})$ such that 
$\forall (\vcxone,\vcxtwo) \in {\cal D}^{(n)}$, 
$(\Phi^{(n)}_{1,\lvckone}(\vcxone),$ $
  \Phi^{(n)}_{2,\lvcktwo}(\vcxtwo))$
$\neq$ $(c_1^{m_1},c_2^{m_2})$. Set 
$$
{\cal B} \defeq \left\{
(\Phi^{(n)}_{1,\lvckone}(\vcxone),
 \Phi^{(n)}_{2,\lvcxtwo}(\vcxtwo)): 
(\vcxone,\vcxtwo)\in {\cal D}^{(n)}
\right\}.
$$  
Then by the above assumption we have
\begin{align}
&{\cal B} \subseteq {\cal X}_1^{m_1} \times {\cal X}_2^{m_2} 
 - \left\{(c_1^{m_1},c_2^{m_2})\right\}.
\label{eqn:SdCCvpp}
\end{align} 
On the other hand we have 
\begin{align*}
&\Psi^{(n)}_{{\lvc k}_1,{\lvc k}_2}({\cal B})
\\
&= \left\{
   \Psi^{(n)}_{{\lvc k}_1,{\lvc k}_2}
  (\Phi^{(n)}_{1,\lvckone}(\vcxone),
   \Phi^{(n)}_{2,\lvcktwo}(\vcxtwo)):
(\vcxone, \vcxtwo)\in {\cal D}^{(n)}
\right\}
\\
&= \left\{
   \psi^{(n)}(\phi^{(n)}_{1}(\vcxone),
         \phi^{(n)}_{2}(\vcxtwo)):
(\vcxone, \vcxtwo)\in {\cal D}^{(n)}
\right\}
={\cal D}^{(n)},
\end{align*}
which together with that 
$\Psi_{{\lvc k}_1,{\lvc k}_2}^{(n)}:$
${\cal X}_1^{m_1}\times {\cal X}_2^{m_2}$$\to$
${\cal X}_1^{n} \times {\cal X}_2^{n}$
is a one-to-one mapping yields that 
\begin{align*}
& |{\cal B}|=|\Psi^{(n)}_{\lvckone, \lvcktwo}({\cal B})|
 =|{\cal D}^{(n)}|=|{\cal X}_1^{m_1}||{\cal X}_2^{m_2}|.
\end{align*} 
The above equality contradicts (\ref{eqn:SdCCvpp}). 
Hence we must have that
$\forall (\vckone,\vcktwo),$
$\forall (c_1^{m_1},c_2^{m_2}),$
$\exists (\vcxone,\vcxtwo)\in {\cal D}^{(n)}$ such that 
$\Phi^{(n)}_{i,{\lvck}_i}(\vcxi)=c_i^{m_i},i=1,2$.
\end{IEEEproof}
}%%%%%%%%%%%%%%%%%%%%%%%%%%%%%%%%%%%%%%%%%%%%%%%%%%
%%%%%%%%%%%%%%%%%%%%%%%%%%%%%%%%%%%%%%%%%%%%%%%%%%%
On the above distributed source encryption scheme, we have 
an important lemma. Before describing this lemma we 
give an observation on 
$p_{C_1^{m_1} C_2^{m_2}|\lrvcxone\lrvcxtwo}$.
For $(\vcxone,\vcxtwo) \in {\cal X}_1^n \times {\cal X}_2^n$, 
we set 
\begin{align*}
&{\cal A}_{\lvcxone,\lvcxtwo}(c_1^{m_1},c_2^{m_2})
\nonumber\\
& \defeq \left\{
(\vckone,\vcktwo): \Phi^{(n)}_{i,\lvcxi}({\vck}_i)= 
c_i^{m_i},i=1,2 
\right\}.
\end{align*}
We have that for each $(c_1^{m_1},c_2^{m_2},\vcxone,\vcxtwo) 
\in {\cal X}_1^{m_1} \times {\cal X}_2^{m_2} %%%%%
\times {\cal X}_1^n \times {\cal X}_2^n$, %%%%%%%%%%%%%%%%%%%%%
\begin{align}%%%%%%%%%%%%%%%%%%%%%%%%%%%%%%%%%%%%%%%%%%%%%%%%%%
&p_{C_1^{m_1} C_2^{m_2}|\lrvcxone\lrvcxtwo}
(c_1^{m_1},c_2^{m_2}|\vcxone,\vcxone) %%%%
\nonumber\\%%%%%%%%%%%%%%%%%%%%%%%%%%%%%%%%%%%%%%%%%%%%%%%%%%%%
&={\rm Pr}\left \{(\rvckone,\rvcktwo)
\in {\cal A}_{\lvcxone,\lvcxtwo}(c_1^{m_1},c_2^{m_2})%%%%%%%%%%%%
\Bigl|\rvcxone=\vcxone,\rvcxtwo=\vcxtwo \right\}%%%%%%%%%%%%%%%%%
\nonumber\\%%%%%%%%%%%%%%%%%%%%%%%%%%%%%%%%%%%%%%%%%%%%%%%%%%%%
&\MEq{a} {\rm Pr}\left\{(\rvckone,\rvcktwo)
 \in {\cal A}_{\lvcxone,\lvcxtwo}(c_1^{m_1},c_2^{m_2})%%%%%% 
\right\}.%%%%%%%%%%%%%%%%%%%%%%%%%%%%
\label{eqn:ddrtrq}
\end{align}
Step (a) follows from $(\rvckone,\rvcktwo) \perp (\rvcxone,\rvcxtwo)$.
We can see from (\ref{eqn:ddrtrq}) that 
for each $(\vcxone, \vcxtwo) \in {\cal X}_1^n \times {\cal X}_2^n$, 
the component 
$p_{C_1^{m_1}C_2^{m_2}|\lrvcxone \lrvcxtwo}
(c_1^{m_1},c_2^{m_2}$ $|\vcxone,\vcxtwo)$ 
of the stochastic matrix 
\begin{align*}
& p_{C_1^{m_1} C_2^{m_2}| \lrvcxone \lrvcxtwo}
(\cdot,\cdot|\vcxone,\vcxtwo) 
\\
&= \left\{ 
 p_{C_1^{m_1} C_2^{m_2}|\lrvcxone, \lrvcxtwo}
   (c_1^{m_1} c_2^{m_2}|\vcxone, \vcxtwo)
   \right\}_{ 
(c_1^{m_1},c_2^{m_2})\in {\cal X}^{m_1} \times {\cal X}^{m_2}}
\end{align*}
can be written as 
\begin{align*}
& p_{C_1^{m_1} C_2^{m_2}| \lrvcxone \lrvcxtwo}
(c_1^{m_1},c_2^{m_2}|\vcxone,\vcxtwo)
\\
&=\Gamma_{\lrvckone \lrvcktwo, (\lvcxone,\lvcxtwo)}(c_1^{m_1},c_2^{m_2}). 
\end{align*}
Furthermore, the quantity 
\begin{align*}
& \Gamma_{\lrvckone \lrvcktwo,(\lvcxone,\lvcxtwo)}
\\
& \defeq \left\{ 
\Gamma_{\lrvckone \lrvcktwo,(\lvcxone,\lvcxtwo)}(c_1^{m_1},c_2^{m_2})
\right \}_{(c_1^{m_1},c_2^{m_2}) 
\in {\cal X}_1^{m_1} \times {\cal X}_2^{m_2}} 
\end{align*}
can be regarded as a joint distribution 
indexed by $(\vcxone, \vcxtwo) \in {\cal X}_1^n \times {\cal X}_2^n$.
Here the random pair 
$(\rvckone,\rvcktwo)$ appearing in 
$\Gamma_{\lrvckone \lrvcktwo,(\lvcxone,\lvcxtwo)}$ 
stands for that the randomness of the joint probability distribution 
is from that of $(\rvckone,\rvcktwo)$. 
From Property \ref{pr:prOne}, we have the following result, which 
is a key result of this paper.    
\begin{lemma}\label{lem:LemOne}
  $\forall (c_1^{m_1},c_2^{m_2}) 
 \in {\cal X}_1^{m_1} \times 
     {\cal X}_2^{m_2}$, we have   
\begin{align*}
\sum_{(\lvcxone, \lvcxtwo) \in {\cal D}^{(n)}}
\Gamma_{\lrvckone \lrvcktwo,(\lvcxone,\lvcxtwo)}
(c_1^{m_1},c_2^{m_2})=1.
%p_{ C_1^{m_1} 
%    C_2^{m_2} | \lrvcxone \lrvcxtwo }
%(c_1^{m_1},c_2^{m_2}| \vcxone, \vcxtwo)=1.
\end{align*}
\end{lemma}

Proof of Lemma \ref{lem:LemOne} is given 
in Appendix \ref{apd:ProofLemOne}. This lemma 
can be regarded as an extension of the Birkhoff-von Neumann
theorem \cite{iwamoto:11}.

\newcommand{\ProofLemOne}{%%%%%%%%%%%%%%%%%%%%%%%%%%%%%%%%%%%%%%%
\subsection{Proof of Lemma \ref{lem:LemOne}}
\label{apd:ProofLemOne}

%}{
In this appendix we prove Lemma \ref{lem:LemOne}.

\begin{IEEEproof}[Proof of Lemma \ref{lem:LemOne}]
By definition we have
\begin{align}
&p_{C_1^{m_1} 
    C_2^{m_2} |     \lrvcxone \lrvcxtwo }
(c_1^{m_1},c_2^{m_2}| \vcxone,  \vcxtwo)
\nonumber\\
&={\rm Pr}\left\{
(\rvckone, \rvcktwo)
\in {\cal A}_{\lvcxone,\lvcxtwo}(c_1^{m_1},c_2^{m_2})
\Bigl| \rvcxone= \vcxone, 
       \rvcxtwo= \vcxtwo 
\right\}
\nonumber\\
&\MEq{a} {\rm Pr}\left\{
(\rvckone, \rvcktwo)
\in {\cal A}_{\lvcxone,\lvcxtwo}(c_1^{m_1},c_2^{m_2})
\right\}.
\label{eqn:ddrtrq}
\end{align}
Step (a) follows from $
      (\rvckone, \rvcktwo) 
\perp (\rvcxone, \rvcxtwo)$.
On the other hand, Property \ref{pr:prOne} part a) 
implies that 
\begin{align}
&     {\cal A}_{\lvcxone, 
                \lvcxtwo}(c_1^{m_1},c_2^{m_2})
 \cap {\cal A}_{{\lvcx}^{\prime}_1,
                {\lvcx}^{\prime}_2}
 (c_1^{m_1},c_2^{m_2})=\emptyset
\notag\\
& \mbox{ for }(\vcxone, 
               \vcxtwo) \neq 
            ({\vcx}^{\prime}_1,
             {\vcx}^{\prime}_2) \in {\cal D}^{(n)}. 
\label{eqn:Serrq}
\end{align}
Furthermore, Property \ref{pr:prOne} part b) implies that  
\begin{align}
 \bigcup_{(\lvcxone,\lvcxtwo) \in {\cal D}^{(n)}}
 {\cal A}_{\lvcxone,\lvcxtwo}(c_1^{m_1},c_2^{m_2})
={\cal X}_1^{n} \times {\cal X}_2^{n}.
\label{eqn:SerrqB}
\end{align}
From (\ref{eqn:ddrtrq}), we have the following chain of 
equalities:
\begin{align*}
& \sum_{(\lvcxone, \lvcxtwo) \in {\cal D}^{(n)}}
 p_{ C_1^{m_1} 
    C_2^{m_2} | \lrvcxone \lrvcxtwo }
 (c_1^{m_1},c_2^{m_2}| \vcxone, \vcxtwo)
\\
&\MEq{a}\Pr \left\{
  (\rvckone, \rvcktwo)
  \in \bigcup_{
  (\lvcxone, \lvcxtwo) \in {\cal D}^{(n)}} 
  {\cal A}_{\lvcxone,\lvcxtwo}(c_1^{m_1},c_2^{m_2})
  \right\}
%\\
%&
\MEq{b}1. 
\end{align*}
Step (a) follows from (\ref{eqn:Serrq}). 
Step (b) follows from (\ref{eqn:SerrqB}).
\end{IEEEproof} 
}%%%%%%%%%%%%%%%%%%%%%%%%%%%%%%%%%%%%%%%%%%%%%%%%%%%%%%%%%%%%%
%%%%%%%%%%%%%%%%%%%%%%%%%%%%%%%%%%%%%%%%%%%%%%%%%%%%%%%%%%%%%%
%%%%%%%%%%%%%%%%%%%%%%%%%%%%%%%%%%%%%%%%%%%%%%%%%%%%%%%%%%%%%%

%From Lemma \ref{lem:LemOne}, we have the following corollary.

\section{%Proposed Security Criterion
Main Results 
}

\subsection{
Proposed Security Criterion
}

In this section, we introduce our proposed security criterion. 
In the following arguments all logarithms are taken 
to the base natural. 
The adversary ${\cal A}$ tries to estimate 
$({\rvcxone},{\rvcxtwo}) \in \mathcal{X}_1 \times \mathcal{X}_1^n$ 
from 
%\begin{align*}
%&
$(C_1^{m_1},$ $C_2^{m_2})$.
%=(\Phi^{(n)}({\rvcx}, {\rvck}), M_{\cal A}^{(n)}) 
%\in \mathcal{X}^{m} \times \mathcal{M}_{\cal A}^{(n)}.
%\end{align*} 
%Note that since $X^n \perp (K^n,Z^n)$, we have 
%$X^n \perp (K^n,$ $M_{\cal A}^{(n)})$. 

The mutual information (MI) between $(\rvcxone,\rvcxtwo)$ and 
$(C_1^{m_1},C_2^{m_2})$ 
denoted by 
$$
\Delta_{\rm MI}^{(n)}
\defeq I(C_1^{m_1}C_2^{m_2};\rvcxone \rvcxtwo)
$$
indicates a leakage of information on $(\rvcxone,\rvcxtwo)$ 
from $(C_1^{m_1},C_2^{m_2})$. 
In this sense it seems to be quite natural 
to adopt the mutual information $\Delta_{\rm MI}^{(n)}$ 
as a security criterion. 
On the other hand, 
directly using $\Delta_{\rm MI}^{(n)}$ as a security criterion 
of the cyptosystem has some problem that this value 
depends on the statistical property of $(\rvcxone,\rvcxtwo)$.
In this paper we propose a new security criterion, which is based on
$\Delta_{\rm MI}^{(n)}$ but overcomes the above problem. 
\begin{definition} 
Let $(\overline{\vc X}_1,\overline{\vc X}_2)$ 
be an arbitrary random variable 
taking values in ${\cal X}_1^n\times {\cal X}_2^n$. 
Set $\overline{C}_i^{m_i}
=\Phi_i^{(n)}({\vc K}_i,\overline{\vc X}_i),i=1,2$.
Define 
\begin{align*}
 {\rm Supp}(p_{\overline{\svc X}_1\overline{\svc X}_2})
  &\defeq \{(\vcxone,\vcxtwo) \in {\cal X}_1^n 
                           \times {\cal X}_2^n:
\\
 &\qquad 
  p_{\overline{\svc X}_1\overline{\svc X}_2}(\vcxone,\vcxtwo) >0\},
\\
 {\cal P}({\cal D}^{(n)})& \defeq \{
p_{\overline{\svc X}_1\overline{\svc X}_2}
\in {\cal P}({\cal X}_1^n \times {\cal X}_2^n):
\\
& \qquad 
{\cal D}^{(n)}
=%\subseteq 
{\rm Supp}(p_{\overline{\svc X}_1\overline{\svc X}_2})=
\}.
\end{align*}
The maximum mutual information criterion denoted 
by $\Delta_{\rm max-MI}^{(n)}$ is as follows. 
\begin{align*}
&\Delta_{{\rm max-MI}}^{(n)}=
 \Delta_{{\rm max-MI}}^{(n)}(
\Phi_1^{(n)},
\Phi_2^{(n)},
\Psi^{(n)}|{p}_{K_1K_2}^n)
 \\
& \defeq 
\max_{p_{\overline{\svc X}_1\overline{\svc X}_2}
\in {\cal P}({\cal D}^{(n)})}
I( \overline{C}_1^{m_1}\overline{C}_2^{m_2}; 
   \overline{\vc X}_1 \overline{\vc X}_2).
\end{align*}
\end{definition} 
Note that in contrast to $\Delta_{\text{MI}}$, 
$\Delta_{\max-\text{MI}}$
does not depend on the distribution of the source. 
Intuitively, one can see $\Delta_{\max-\text{MI}}$ as a 
metric similar to channel capacity. We further define the 
following quantity. 
%Define 
\begin{align*}
&\overline{\Delta}_{{\rm max-MI}}^{(n)}
=\overline{\Delta}_{{\rm max-MI}}^{(n)}( 
\Phi_1^{(n)},
\Phi_2^{(n)}|{p}_{K_1K_2}^n)
\\
& \defeq 
\max_{p_{\overline{\svc X}_1\overline{\svc X}_2}
\in {\cal P}({\cal X}_1^n \times {\cal X}_2^n)}
I(\overline{C}_1^{m_1}
  \overline{C}_2^{m_2}; 
  \overline{\vc X}_1 
  \overline{\vc X}_2).
\end{align*}

By definition it is obvious that 
$\Delta_{\rm MI}^{(n)} \leq 
\overline{\Delta}_{{\rm max-MI}}^{(n)}$
and 
$\Delta_{\rm max-MI}^{(n)} \leq 
\overline{\Delta}_{{\rm max-MI}}^{(n)}$.
We have the following proposition
on $\Delta_{{\rm MI}}^{(n)}$,
$\Delta_{{\rm max-MI}}^{(n)}$, and 
$\overline{\Delta}_{{\rm max-MI}}^{(n)}$: 
%%%%%%%%%%%%%%%%%%%%%%%%%%%%%%%%%%%%%%%%%%%%%%%%%%%%%%%%%
%which %%%%%%%%%%%%%%%%%%%%%%%%%%%%%%%%%%%%%%%%%%%%%%%%%%
%is the most essential part in the proof of the %%%%%%%%%
%strong converse theorem. %%%%%%%%%%%%%%%%%%%%%%%%%%%%%%%
%%%%%%%%%%%%%%%%%%%%%%%%%%%%%%%%%%%%%%%%%%%%%%%%%%%%%%%%%
\begin{proposition}\label{pro:ProOneA}
$\quad$
\begin{itemize}
\item[a)] We have the following:
$$
\max\{{\Delta}_{{\rm MI}}^{(n)},
      {\Delta}_{{\rm max-MI}}^{(n)}
\}\leq \overline{\Delta}_{{\rm max-MI}}^{(n)}.
$$
\item[b)] We assume that 
$
{\cal D}^{(n)} \subseteq {\rm Supp}(p_{X_1X_2}^n)
={\rm Supp}^n(p_{X_1X_2}).
$
Under this assumption, if 
$\Delta_{\rm MI}^{(n)}=$ $I(C_1^{m_1}C_2^{m_2};\rvcxone\rvcxtwo)=0,$ 
then we have $\Delta_{{\rm max-MI}}^{(n)}=0$. 
This implies that $\Delta_{{\rm max-MI}}^{(n)}$ is valid as a 
measure of information leakage. 
\item[c)] We have the following.
\begin{align*}
& {\Delta}_{{\rm max-MI}}^{(n)} 
\\
& \geq \ba[t]{ll}
\max \{& m_1\log{|{\cal X}_1|}-nH(K_1), \\
       & m_2\log{|{\cal X}_2|}-nH(K_2), \\
       & m_1\log{|{\cal X}_1|}
        +m_2\log{|{\cal X}_2|}-nH(K_1K_2)\}.
    \ea
\end{align*}
\end{itemize}
\end{proposition}

Proof of Proposition \ref{pro:ProOneA} is given in 
Appendix \ref{apd:ProofProOneA}.
%%%%%%%%%%%%%%%%%%%%%%%%%%%%%%%%%%%%%%%%%%%%%%%%%%%%%%%%%%%%%%%%%% 
%The part b) of this lemma can be regarded as an extension of the 
%Birkhoff-von Neumann theorem \cite{iwamoto:11}.
%%%%%%%%%%%%%%%%%%%%%%%%%%%%%%%%%%%%%%%%%%%%%%%%%%%%%%%%%%%%%%%%%
\newcommand{\ProofProOneA}{%%%%%%%%%%%%%%%%%%%%%%%%%%%%%%%%%%%%%%%
\subsection{Proof of Proposition \ref{pro:ProOneA}}
\label{apd:ProofProOneA}%%%%%%%%%%%%%%%%%%%%%%%%%%%%%%%%%%%%%%%%%%
%%%%%%%%%%%%%%%%%%%%%%%%%%%%%%%%%%%%%%%%%%%%%%%%%%%%%%%%%%%%%%%%%
%%%%%%%%%%%%%%%%%%%%%%%%%%%%%%%%%%%%%%%%%%%%%%%%%%%%%%%%%%%%%%%%%
%In this appendix we prove Proposition \ref{pro:ProOne}.%%%%%%%%%
%%%%%%%%%%%%%%%%%%%%%%%%%%%%%%%%%%%%%%%%%%%%%%%%%%%%%%%%%%%%%%%%%
%}{

In this appendix we prove Proposition \ref{pro:ProOneA}. 

\begin{IEEEproof}[Proof of Proposition \ref{pro:ProOneA}] 
The part a) is obvious. We first prove the part b). Using the quntities 
\begin{align*} 
& \Gamma_{\lrvckone \lrvcktwo,(\lvcxone,\lvcxtwo)}(c_1^{m_1},c_2^{m_2}), 
\\
& (\vcxone,\vcxtwo, c_1^{m_1},c_1^{m_1}) 
   \in {\cal X}_1^n   \times {\cal X}_2^n 
\times {\cal X}^{m_1} \times {\cal X}^{m_1}, 
\end{align*} 
components $p_{C_1^{m_1}C_2^{m_2}|\lrvcxone \lrvcxtwo}
(c_1^{m_1},c_2^{m_2})$ of the joint distribution 
$p_{C_1^{m_1}C_2^{m_2}}$ can be computed as 
%%%%%%%%%%%%%%%%%%%%%%%%%%%%%%%%%%%%%%%%%%%%%%%%%%%%%%%%%%
%%%%%%%%%%%%%%%%%%%%%%%%%%%%%%%%%%%%%%%%%%%%%%%%%%%%%%%%%%
%$(c^m,a) \in {\cal X}^m \times {\cal M}_{\cal A}^{(n)}$%%
%%%%%%%%%%%%%%%%%%%%%%%%%%%%%%%%%%%%%%%%%%%%%%%%%%%%%%%%%%
%%%%%%%%%%%%%%%%%%%%%%%%%%%%%%%%%%%%%%%%%%%%%%%%%%%%%%%%%%
\begin{align*}
& p_{C_1^{m_1}C_2^{m_2}}(c_1^{m_1},c_2^{m_2})
\\
&= \sum_{(\lvcxone,\lvcxtwo)} p_{\lrvcxone\lrvcxtwo}
(\vcxone, \vcxtwo)
\Gamma_{\lrvckone \lrvcktwo,(\lvcxone, \lvcxtwo)}(c_1^{m_1}c_2^{m_2}). 
\end{align*}
Set 
\begin{align*}
&\Gamma_{\lrvckone \lrvcktwo}
 ^{(p_{\srvcxone \srvcxtwo})}(c_1^{m_1},c_2^{m_2})
\\
& =\sum_{(\lvcxone,\lvcxtwo)}p_{\lrvcxone \lrvcxtwo}(\vcxone, \vcxtwo)
  \Gamma_{\lrvckone \lrvcktwo,(\lvcxone,\lvcxtwo)}(c_1^{m_1},c_2^{m_2})
\\
&= p_{C_1^{m_1}C_2^{m_2}}(c_1^{m_1},c_2^{m_2}).
\end{align*}
Furthermore, set 
\begin{align*}
& \Gamma^{(p_{\srvcxone \srvcxtwo})}_{\lrvckone\lrvcktwo}
\defeq \left\{
\Gamma_{\lrvckone \lrvcktwo }^{(p_{\srvcxone\srvcxtwo})}
(c_1^{m_1},c_2^{m_2}) \right\}_{(c_1^{m_1}, c_2^{m_2}) 
\in {\cal X}_1^{m_1} \times {\cal X}_2^{m_2} }
\\
&=p_{C_1^{m_1}C_2^{m_2}}.
\end{align*}
Using 
$\Gamma_{\lrvckone\lrvckone,(\lvcxone,\lvcxtwo)}, 
     (\vcxone,\vcxtwo) 
\in {\cal X}_1^n\times {\cal X}_2^n$ 
and $\Gamma_{\lrvckone\lrvcktwo}^{(p_{\srvcxone\srvcxtwo})}$,    
we compute $\Delta_{\rm MI}^{(n)}$ to obtain 
\begin{align}
&\Delta_{\rm MI}^{(n)}=I(C_1^{m_1}C_2^{m_2}; \rvcxone \rvcxtwo )
 =\sum_{\scs (\lvcxone,\lvcxtwo) 
 \atop{ \scs \in {\cal X}_1^n \times {\cal X}_2^n}}
p_{\lrvcxone\lrvcxtwo}(\vcxone,\vcxtwo)
\notag\\
&\quad \qquad \times 
D\left(
\Gamma_{\lrvckone\lrvcktwo,(\lvcxone,\lvcxtwo)} \Big|\Big| 
\Gamma^{(p_{\srvcxone\srvcxtwo})}_{\lrvckone\lrvcktwo}\right)
\notag\\
& \geq \sum_{\scs (\lvcxone,\lvcxtwo)\in {\cal D}^{(n)}}
p_{\lrvcxone\lrvcxtwo}(\vcxone,\vcxtwo)
\notag\\
&\quad \qquad \times 
D\left(
\Gamma_{\lrvckone\lrvcktwo,(\lvcxone,\lvcxtwo)} \Big|\Big| 
\Gamma^{(p_{\srvcxone\srvcxtwo})}_{\lrvckone\lrvcktwo}\right).
\label{eqn:SDDxx}
\end{align}
%We note that 
%since $(\rvcxone,\rvcxtwo)$ 
%is from the discrete memolyless 
%source specified with $p_{X_1X_2}$, 
By the assumption 
${\cal D}^{(n)}\subseteq {\rm Supp}^n(p_{X_1X_2})$, 
we have that 
\begin{align}
&p_{\lrvcxone\lrvcxtwo}(\vcxone,\vcxtwo) 
= \prod_{t=1}^n p_{X_1X_2}(x_{1,t},x_{2,t})>0, 
\notag\\
& \forall (\vcxone,\vcxtwo) 
\in {\cal D}^{(n)}. %{\cal X}_1^n \times {\cal X}_2^n.
\label{eqn:Sffxx}
\end{align}
Now we suppose that $\Delta_{\rm MI}^{(n)}=0$. 
%I(C^{m}; \rvcx | { }x)=0$. 
Then from (\ref{eqn:SDDxx}) and (\ref{eqn:Sffxx}), we have 
\begin{align}
& \Gamma_{\lrvckone\lrvcktwo, (\lvcxone,\lvcxtwo)}
= \Gamma_{\lrvckone\lrvcktwo, ({\lvcx}_1^{\ast},{\lvcx}_2^{\ast})}
= \Gamma^{(p_{\srvcxone\srvcxtwo})}_{\lrvckone\lrvcktwo}, 
\notag\\
& \forall (\vcxone,\vcxtwo) \in {\cal D}^{(n)},
\label{eqn:ssSDDxppx}
\end{align}
where $({\vcx}_1^{\ast},{\vcx}_2^{\ast})$ is an 
element of ${\cal D}^{(n)}$. 
Let $(\overline{\vc X}_{1,{\rm opt}},
      \overline{\vc X}_{2,{\rm opt}})$ 
be the optimal random variable, the distribution 
$p_{\overline{\lvc X}_{1,{\rm opt}}
    \overline{\lvc X}_{2,{\rm opt}}}$ of which 
attains the maximum in the definition of 
$\Delta_{{\rm max-MI}}^{(n)}$. 
We set 
$\overline{C}^m_{i,{\rm opt}}
=\Phi^{(n)}_i({\vc K}_i,\overline{\vc X}_{i,{\rm opt}})$,
$i=1,2$. By definition we have 
$$
\Delta_{{\rm max-MI}}^{(n)}
=I(\overline{C}_{1,\rm opt}^{m_1}
   \overline{C}_{2,\rm opt}^{m_2}; 
\overline{\rvcx}_{1,\rm opt}
\overline{\rvcx}_{2,\rm opt}^n).
$$ 
Using (\ref{eqn:ssSDDxppx}), we compute 
$\Gamma^{(p_{\overline{\srvcx}_{1,\rm opt}
             \overline{\srvcx}_{2,\rm opt}})}
_{\lrvckone \lrvcktwo}(c_1^{m_1},c_2^{m_2})$, 
$(c_1^{m_1},$ $c_2^{m_2}) 
\in {\cal X}_1^{m_1} \times {\cal X}_2^{m_2}$ 
to obtain
\begin{align*}
&\Gamma^{(p_{\overline{\srvcx}_{1,\rm opt}
             \overline{\srvcx}_{2,\rm opt}})}
_{\lrvckone \lrvcktwo}(c_1^{m_1},c_2^{m_2})
\\
&=\sum_{\scs (\lvcxone,\lvcxtwo)\in {\cal D}^{(n)}}
p_{\overline{\lrvcx}_{1,\rm opt}
   \overline{\lrvcx}_{2,\rm opt}}(\vcxone,\vcxtwo)
\notag\\
&\quad \times 
\Gamma_{\lrvckone\lrvcktwo,(\lvcxone,\lvcxtwo)}
(c_1^{m_1},c_2^{m_2})
 =\Gamma_{\lrvckone\lrvcktwo,
({\lvcx}_1^{\ast},{\lvcx}_2^{\ast})}(c_1^{m_1},c_2^{m_2}).
\end{align*}
Hence we have
\begin{align}
&\Gamma^{(p_{\overline{\srvcx}_{1,\rm opt}
             \overline{\srvcx}_{2,\rm opt}})}
_{\lrvckone \lrvcktwo}(c_1^{m_1},c_2^{m_2})
=\Gamma_{\lrvckone\lrvcktwo,
 ({\lvcx}_1^{\ast},{\lvcx}_2^{\ast})}(c_1^{m_1},c_2^{m_2})
\notag\\
&=\Gamma_{\lrvckone\lrvcktwo,(\lvcxone,\lvcxtwo)}
(c_1^{m_1},c_2^{m_2}),
\forall (\vcxone, \vcxtwo) \in {\cal D}^{(n)}.
\label{eqn:jjDDxppx}
\end{align}
From (\ref{eqn:jjDDxppx}), we have
\begin{align*}
&\Delta_{{\rm max-MI}}^{(n)}
=I(\overline{C}_{1,\rm opt}^m 
   \overline{C}_{2,\rm opt}^m;
   \overline{\rvcx}_{1,\rm opt} 
   \overline{\rvcx}_{2,\rm opt})
\\
& =\sum_{\scs (\lvcxone,\lvcxtwo) 
         \scs \in {\cal D}^{(n)}}
p_{\overline{\lrvcx}_{1,\rm opt} 
   \overline{\lrvcx}_{2,\rm opt}}(\vcxone,\vcxtwo)
\\
& \quad \qquad \times 
D\left(
\Gamma_{\lrvckone \lrvcktwo,(\lvcxone,\lvcxtwo)}
\Big|\Big| \Gamma^{(p_{\overline{\srvcx}_{1,\rm opt}
                       \overline{\srvcx}_{2,\rm opt}})
          }_{\lrvckone\lrvcktwo}
\right)=0.
\end{align*}
We next prove the part c).
%\begin{figure}[t]
%\centering
%\includegraphics[width=0.28\textwidth]{CheckRVEncGnlb.eps}
%\caption{$\crvcx$, $\crvcc$ and ${ }$.
%\label{fig:CheckRV}}
%\end{figure}
Let $(\crvcxone, \crvcxtwo)$ be a pair 
of uniformly distributed random vectors  
over ${\cal D}^{(n)}$. Set $\check{C}_i^{m_i} 
\defeq$ $\Phi_{i,\lrvcki}(\crvcxi)$, 
$i=1,2$.
%%%%%%%%%%%%%%%%%%%%%%%%%%%%%%%%%%%%%%%%%
%The three random variables $\crvcx$, 
%$\check{C}^{m}$vare shown 
%in Fig. \ref{fig:CheckRV}. 
We claim that 
$(\check{C}_1^{m_2},\check{C}_2^{m_2})$ 
is the uniformly distributed random pair
over ${\cal X}_1^{m_1}\times {\cal X}_2^{m_2}$. 
In fact for each 
$(c_1^{m_1},c_2^{m_2}) \in 
{\cal X}_1^{m_1}\times {\cal X}_2^{m_2}$, 
we have the following chain of equalities:
%%%%%%%%%%%%%%%%%%%%%%%%%%%%%%%%%%%%%%%%%%%%%%%%%%%%%%%%%%%%%%
%%%%%%%%%%%%%%%%%%%%%%%%%%%%%%%%%%%%%%%%%%%%%%%%%%%%%%%%%%%%%%
\begin{align}
& |{\cal X}_1^{m_1}||{\cal X}_2^{m_2}|
   p_{\check{C}_1^{m_1}\check{C}_2^{m_2}}
(c_1^{m_1},c_2^{m_2}) \MEq{a}|{\cal D}^{(n)}| 
\sum_{\scs (\lvcxone,\lvcxtwo) \in {\cal D}^{(n)}}1 
\notag\\
&\quad \times p_{\check{C}_1^{(m_1)}\check{C}_2^{(m_2)} 
 |\lcrvcxone \lcrvcxtwo}(c_1^{m_1},c_2^{m_2}|\vcxone,\vcxtwo) 
\cdot\frac{1}{|{\cal D}^{(n)}|}
\notag\\
&=\sum_{\scs (\lvcxone, \lvcxtwo){\scs \in {\cal D}^{(n)}}}
\Gamma_{\lrvckone \lrvcktwo,(\lvcxone,\lvcxtwo)}(c_1^{m_1},c_2^{m_2})
\MEq{b}1. 
\label{eqn:SDfffp}
\end{align}
Step (a) follows from Property \ref{pr:prOnDecSet}.
Step (b) follows from Lemma \ref{lem:LemOne}.
Since we have (\ref{eqn:SDfffp}) for every 
$(c_1^{m_1},c_2^{m_2}) 
\in {\cal X}_1^{m_1} \times {\cal X}_2^{m_2} $, 
we have that $(\check{C}_1^{m_1},\check{C}_2^{m_2})$ 
is the uniformly distributed 
random pair over ${\cal X}_1^{m_1} 
\times {\cal X}_2^{m_2}$. 
We have the following chain of inequalities:
%%%%%%%%%%%%%%%%%%%%%%%%%%%%%%%%%%%%%%%%%%%%%%%%%%%%%%%%%%%%%%%%
%%%%%%%%%%%%%%%%%%%%%%%%%%%%%%%%%%%%%%%%%%%%%%%%%%%%%%%%%%%%%%%%
\begin{align*}
& \Delta_{{\rm max-MI}}^{(n)}\geq 
I(\check{C}_1^{m_1}\check{C}_2^{m_2};\crvcxone \crvcxtwo)
\nonumber\\ 
&= H(\check{C}_1^{m_1}\check{C}_2^{m_2}) 
  -H(\check{C}_1^{m_1}\check{C}_2^{m_2}|\crvcxone\crvcxtwo)
\nonumber\\ 
&\MEq{a} m_1 \log |{\cal X}_1| +m_2 \log |{\cal X}_2|
 -H(\check{C}_1^{m_1}\check{C}_1^{m_1}|\crvcxone\crvcxtwo).
\\
&=m_1\log |{\cal X}_1| + m_2\log |{\cal X}_2|
\\
&\quad -H(\Phi_1^{(n)}(\rvckone,\crvcxone)
          \Phi_2^{(n)}(\rvckone,\crvcxone)
          |\crvcxone \crvcxtwo)
\\
& \MGeq{b} 
m_1 \log{|{\cal X}_1|}+m_2\log{|{\cal X}_2|}
-H(\rvckone\rvcktwo|\crvcxone\crvcxtwo)
\\
&=m_1\log{|{\cal X}_1|}+m_2\log{|{\cal X}_2|}-nH(K_1K_2).
\end{align*}
Step (a) follows from 
that 
$(\check{C}_1^{m_2},\check{C}_2^{m_2})$ 
is the uniformly distributed random pair
over ${\cal X}_1^{m_1}\times {\cal X}_2^{m_2}$. 
Step (b) follows from the data processing inequality.
Furthermore for $i=1,2$, we have the following chain 
of inequalities:
%%%%%%%%%%%%%%%%%%%%%%%%%%%%%%%%%%%%%%%%%%%%%%%%%%%%%%%%%%%%%%%%
%%%%%%%%%%%%%%%%%%%%%%%%%%%%%%%%%%%%%%%%%%%%%%%%%%%%%%%%%%%%%%%%
\begin{align*}
& \Delta_{{\rm max-MI}}^{(n)}\geq 
I(\check{C}_1^{m_1}\check{C}_2^{m_2};\crvcxone \crvcxtwo)
\geq 
I(\check{C}_i^{m_i};\crvcxi)
\nonumber\\ 
&= H(\check{C}_i^{m_i}) -H(\check{C}_i^{m_i}| \crvcxi)
%\nonumber\\ 
%&
\MEq{a} m_i \log |{\cal X}_i| 
 -H(\check{C}_i^{m_i}|\crvcxi)
\\
&=m_i\log |{\cal X}_i|-H(\Phi_i^{(n)}(\rvcki,\crvcxi)|\crvcxi)
\\
& \MGeq{b} m_i \log{|{\cal X}_i|}-H(\rvcki|\crvcxi)
   =m_i\log{|{\cal X}_i|}-nH(K_i).
\end{align*}
Step (a) follows from 
that for $i=1,2$,
$\check{C}_i^{m_i}$ 
is the uniformly distributed random variable 
over ${\cal X}_i^{m_i}$. 
Step (b) follows from the data processing inequality.
\end{IEEEproof}
\noindent
}%%%%%%%%%%%%%%%%%%%%%%%%%%%%%%%%%%%%%%%%%%%%%%%%%%%%%%%%%%%
%%%%%%%%%%%%%%%%%%%%%%%%%%%%%%%%%%%%%%%%%%%%%%%%%%%%%%%%%%%%
%%%%%%%%%%%%%%%%%%%%%%%%%%%%%%%%%%%%%%%%%%%%%%%%%%%%%%%%%%%%
%%%%%%%%%%%%%%%%%%%%%%%%%%%%%%%%%%%%%%%%%%%%%%%%%%%%%%%%%%%%
%%%%%%%%%%%%%%%%%%%%%%%%%%%%%%%%%%%%%%%%%%%%%%%%%%%%%%%%%%%%
%%%%%%%%%%%%%%%%%%%%%%%%%%%%%%%%%%%%%%%%%%%%%%%%%%%%%%%%%%%%
%%%%%%%%%%%%%%%%%%%%%%%%%%%%%%%%%%%%%%%%%%%%%%%%%%%%%%%%%%%%
%%%%%%%%%%%%%%%%%%%%%%%%%%%%%%%%%%%%%%%%%%%%%%%%%%%%%%%%%%%%
%%%%%%%%%%%%%%%%%%%%%%%%%%%%%%%%%%%%%%%%%%%%%%%%%%%%%%%%%%%%
%%%%%%%%%%%%%%%%%%%%%%%%%%%%%%%%%%%%%%%%%%%%%%%%%%%%%%%%%%%%
%%%%%%%%%%%%%%%%%%%%%%%%%%%%%%%%%%%%%%%%%%%%%%%%%%%%%%%%%%%%
%%%%%%%%%%%%%%%%%%%%%%%%%%%%%%%%%%%%%%%%%%%%%%%%%%%%%%%%%%%%
%%%%%%%%%%%%%%%%%%%%%%%%%%%%%%%%%%%%%%%%%%%%%%%%%%%%%%%%%%%%
%%%%%%%%%%%%%%%%%%%%%%%%%%%%%%%%%%%%%%%%%%%%%%%%%%%%%%%%%%%%
%%%%%%%%%%%%%%%%%%%%%%%%%%%%%%%%%%%%%%%%%%%%%%%%%%%%%%%%%%%%
%%%%%%%%%%%%%%%%%%%%%%%%%%%%%%%%%%%%%%%%%%%%%%%%%%%%%%%%%%%%
%%%%%%%%%%%%%%%%%%%%%%%%%%%%%%%%%%%%%%%%%%%%%%%%%%%%%%%%%%%%
\begin{remark}
The part b) in the above proposition is quite essential. If we have 
a security criterion $\widehat{\Delta}^{(n)}$ not satisfying this 
condition, it may happen that 
$
\Delta_{\rm MI}^{(n)}=I(C_1^{m_1}C_2^{m_2};
\rvcxone \rvcxtwo)=0, 
$ 
but $\widehat{\Delta}^{(n)}>0$. Such $\widehat{\Delta}^{(n)}$ 
is invalid for the security criterion. 
\end{remark}
\begin{remark}
The property stated in the part c) is a key important property 
of $\Delta_{\rm max-MI}^{(n)}$, which plays an important role 
in establishing the strong converse theorem. Lemma \ref{lem:LemOne} 
is a key result for the proof of the part b).
\end{remark}

%%%%%%%%%%%%%%%%%%%%%%%%%%%%%%%%%%%%%%%%%%%%%%%%%%%%%%%%%%%%%
\newcommand{\OmittZZ}{%%%%%%%%%%%%%%%%%%%%%%%%%%%%%%%%%%%%%%%
Let $(\check{C}_1^{m_1},
      \check{C}_2^{m_2},
      \check{\rvcx}_1^n, 
      \check{\rvcx}_2^n)$ be a quadruple of random variables. 
We assume that
$ p_{\check{C}_1^{m_1}\check{C}_2^{m_2}|
   \check{\lrvcx}_1 \check{\lrvcx}_2}
  =p_{C_1^{m_1}C_2^{m_2}|\lrvcxone 
                        \lrvcxtwo}$.
We further assume that 
$ p_{\check{\lrvcx}_1 
     \check{\lrvcx}_2}$ 
is the uniform distribution over ${\cal D}^{(n)}$.
Then by Lemma \ref{lem:LemOne} we have that
\begin{align*}
& \sum_{\scs (\lvcxone, \lvcxtwo) 
        %\atop
        {\scs \in {\cal D}^{(n)}}}
\hspace*{-3mm}
p_{\check{C}_1^{m_1}\check{C}_2^{m_2} 
   \check{\lrvcx}_1 
   \check{\lrvcx}_2}
(c_1^{m_1},c_2^{m_2}, \vcxone, \vcxtwo) 
=\frac{1}{|{\cal X}_1^{m_1}||{\cal X}_2^{m_2}|}.
\end{align*}
Hence $p_{\check{C}_1^{m_1} \check{C}_2^{m_2}}$
is the uniform distribution over 
${\cal X}_1^{m_1} \times{\cal X}_2^{m_2}$. 
}%%%%%%%%%%%%%%%%%%%%%%%%%%%%%%%%%%%%%%%%%%%%%%%%%%%%%%%%%
%%%%%%%%%%%%%%%%%%%%%%%%%%%%%%%%%%%%%%%%%%%%%%%%%%%%%%%%%%
%Form Lemma \ref{lem:LemOne} we have the following result.
%\begin{corollary}
%For any $(\lrvcxone, \lrvcxtwo)$ and its encrytion
%$({C}_1^{m_1},{C}_2^{m_2})$, we have that if 
%$
%I(C_1^{m_1}C_2^{m_2}; \rvcxone \rvcxtwo)=0, then
%$
%the following
%\end{corollary}
%For any distributed source encryption 

%%%%%%%%%%%%%%%%%%%%%%%%%%%%%%%%%%%%%%%%%%%%%%%%%%%%%%
%%%%%%%%%%%%%%%%%%%%%%%%%%%%%%%%%%%%%%%%%%%%%%%%%%%%%%

\underline{\it Defining Reliability and Security:} 
%From the common key cryptosystem 
%shown in Fig. \ref{fig:main},
The decoding process is successful if 
$(\widehat{\rvcx}_1, 
  \widehat{\rvcx}_2)=(\rvcxone,$ $
                 \rvcxtwo)$ holds.
%%%%%%%%%%%%%%%%%%%%%%%%%%%%%%%%%%%%%%%%%%%%%%%%%%%%%%%%%%%%%%%%%%%%
%Combining this and (\ref{eq:source_estimation}), %%%%%%%%%%%%%%%%%%
%%%%%%%%%%%%%%%%%%%%%%%%%%%%%%%%%%%%%%%%%%%%%%%%%%%%%%%%%%%%%%%%%%%%
Hence the decoding error probability 
is given by  
\begin{align*}
&\Pr[\Psi^{(n)}(\rvckone, \rvcktwo,
  \phi_1^{(n)}({\rvckone, \rvcxone}),
  \phi_2^{(n)}({\rvcktwo, \rvcxtwo}))
\\
&\qquad \neq ({\rvcxone},{\rvcxtwo}) ]
\\
&=\Pr[\Psi^{(n)}_{\lrvckone,\lrvcktwo}
(\Phi_{1,\lrvckone}^{(n)}(\rvcxone),
 \Phi_{2,\lrvcktwo}^{(n)}(\rvcxtwo))
\neq ({\rvcxone},{\rvcxtwo}) ]
\\
&=\Pr[\psi^{(n)}
(\phi_{1}^{(n)}(\rvcxone),
 \phi_{2}^{(n)}(\rvcxtwo))
\neq           (\rvcxone,
                \rvcxtwo)]
\\
&=\Pr[(\rvcxone,\rvcxtwo)\notin {\cal D}^{(n)}].
\end{align*}
Since the above quantity depends only on 
$(\phi_1^{(n)}, \phi_2^{(n)},\psi^{(n)})$,
we wirte the error probability $p_{{\rm e}}$ of decoding as 
\begin{align*}
p_{{\rm e}}=&p_{{\rm e}} 
(\phi_1^{(n)},\phi_2^{(n)},\psi^{(n)}
|{p}_{X_1X_2}^n,{p}_{K_1K_2}^n)
%\\
%p_{{\rm e}}(\varphi^{(n)},\psi^{(n)}|{p}_{X_1X_2}^n,{p}_{K_1K_2}^n)
%p_{{\rm e}}(\varphi^{(n)},\psi^{(n)}|{p}_{X_1X_2}^n,{p}_{K_1K_2}^n)
\\ 
\defeq & \Pr[(\rvcxone,\rvcxtwo) \notin {\cal D}^{(n)})].
\end{align*}
%Since $\Delta^{(n)}$ depends on 
%$(\Phi_1^{(n)},\Phi_2^{(n)}, \Psi^{(n)})$, we write this quantity 
%as 
%$$
%\Delta^{(n)}=\Delta^{(n)}(
%\Phi_1^{(n)},
%\Phi_2^{(n)},
%\Psi^{(n)}|{p}_{X_1X_2}^n,{p}_{K_1K_2}^n).
%$$
\begin{definition} We fix some positive constant $\varepsilon_0$. 
        For a fixed pair $(\varepsilon, \delta) 
        \in [0,\varepsilon_0] \times (0,1)$, 
        $(R_1,R_2)$ is $(\varepsilon,\delta)$-admissible 
        if there exists a sequence 
        $\{(\Phi_1^{(n)},\Phi_2^{(n)},$ $\Psi^{(n)})\}_{n \geq 1}$
        such that $\forall \gamma >0$,
        $\exists n_0=n_0(\gamma) \in \mathbb{N}$, 
	$\forall n\geq n_0$, we have 
	\begin{align*}
        &\frac{1}{n} \log |{\cal X}_i^{m_i}| 
         = \frac{m_i}{n} \log |{\cal X}_i| \in 
         \left[R_i-\gamma, R_i+ \gamma \right],i=1,2,
\\  				
& p_{{\rm e}}(
\phi_1^{(n)},
\phi_2^{(n)},
\psi^{(n)}|{p}_{X_1X_2}^n,{p}_{K_1K_2}^n)
\leq \delta,
\\
&%\mbox{ and } 
\Delta^{(n)}_{\rm max-MI}(
\Phi_1^{(n)},
\Phi_2^{(n)},
\Psi^{(n)}|{p}_{K_1K_2}^n) 
\leq \varepsilon.
\end{align*}
\end{definition}
\begin{definition}{\bf (Reliable and Secure Rate Set)}
        Let $\mathcal{R}(\varepsilon,$ $\delta|p_{X_1X_2},$ $p_{K_1K_2})$
	denote the set of all $(R_1,R_2)$ such that $(R_1,$ $R_2)$ 
        is 
        $(\varepsilon,\delta)$-admissible. Furthermore, set 
        $$
         \mathcal{R}(p_{X_1X_1},p_{K_1K_2}) \defeq  
        \bigcap_{\scs (\varepsilon, \delta) \in (0,\varepsilon_0] 
                 \atop{\scs \times (0,1)
                 }
         }
\mathcal{R}(\varepsilon,\delta|
p_{X_1X_2},
p_{K_1K_2})
        $$
We call $\mathcal{R}(
p_{X_1X_2},
p_{K_1K_2})$ 
       the \emph{\bf reliable and secure rate} set. 
\end{definition}

\subsection{
Strong Converse for the Distributed Source Encryption
}

To state our results on $\mathcal{R}(\varepsilon,\delta|
p_{X_1X_2},
p_{K_1K_2})$ for $(\varepsilon,\delta) 
            \in [0,\varepsilon_0] \times (0,1)$, 
define the following two regions:
\begin{align*}
 \mathcal{R}_{\mathrm{sw}}(p_{X_1X_2}):=\{(R_1,R_2) : 
%\\
     \ &R_1 \geq H(X_1|X_2), 
\\
& R_2 \geq H(X_2|X_1), 
\\
& 
  R_1+R_2 \geq H(X_1X_2)\},
\\
\mathcal{R}_{\mathrm{key}}(p_{K_1K_2}):=\{(R_1,R_2): 
%\\
\ &R_1 \leq H(K_1),R_2 \leq H(K_2),
\\
& R_1+R_2 \leq  H(K_1K_2)\}.
\end{align*}

Santoso and Oohama 
\cite{DBLP:conf/isit/SantosoO17}, 
\cite{santosoOhPEC:19}
proved that the bound 
$\mathcal{R}_{\mathrm{key}}(p_{K_1K_2})$
$\cap$ 
$\mathcal{R}_{\mathrm{sw}}(p_{X_1X_2})$
serves as  an inner bound of 
$\mathcal{R}(p_{X_1X_2}$, 
$p_{K_1K_2})$ in the case where the security criterion 
is measured by the mutual information 
$\Delta_{\rm MI}^{(n)}$. By a simple observation we can 
see that their post encryption compression scheme yields 
the same bound in the present case of security criterion 
mesured by $\Delta_{\rm \max-MI}^{(n)}$. Hence we have the 
following theorem:  
\begin{theorem}
\label{th:directTh}
For each $(\varepsilon, \delta) 
\in (0,\varepsilon_0] \times (0,1)$, we have
\begin{align}
&      \mathcal{R}_{\mathrm{key}}(p_{K_1K_2})
  \cap \mathcal{R}_{\mathrm {sw}}(p_{X_1X_2})
\notag\\
&\subseteq \mathcal{R}(p_{X_1X_2},p_{K_1K_2})  
 \subseteq \mathcal{R}(\varepsilon,\delta|
                      p_{X_1X_2},p_{K_1K_2}).
\notag
\end{align} 
\end{theorem}

Outline of the proof of this theorem will be given in the 
next section. We next derive one outer bound by a simple 
observation based on previous works on the distributed 
source coding for correlted sources. 
%%%%%%%%%%%%%%%%%%%%%%%%%%%%%%%%%%%%%%%%%%%%%%%%%%%%%%
From the communication 
scheme %shown in Fig. \ref{fig:main}, 
%%%%%%%%%%%%%%%%%%%%%%%%%%%%%%%%%%%%%%%%%%%%%%%%%%%%
we can see that the common key cryptosysytem 
can be regarded as the data compression system, where 
for each $i=1,2$, the encoder $\Phi_i^{(n)}$ and the decoder 
$\Psi^{(n)}$ can use the common side information $\rvcki$.
By the strong converse coding theorem for 
this data compression system \cite{oohama:94}, we have that if 
\begin{align*}
& R_1 < H(X_1|X_2K_1K_2)=H(X_1|X_2) \mbox{ or }
\\
& R_2 < H(X_2|X_1K_1K_2)=H(X_2|X_1) \mbox{ or }
\\
& R_1+ R_2 < H(X_1X_2|K_1K_2)=H(X_1X_2) 
\end{align*}
then 
$\forall \tau \in (0,1)$, $\forall \gamma >0$,
and $\forall \{(\phi_1^{(n)},\phi_2^{(n)},$ $\psi^{(n)})\}_{n \geq 1}$,
        $\exists n_0=n_0(\tau,\gamma) \in \mathbb{N}$, 
	$\forall n\geq n_0$, we have the following:
	\begin{align*}
        & \frac{m}{n} \log |{\cal X}_i|\leq R_i + \gamma, i=1,2,\:
        \\
        & p_{{\rm e}}(\phi_1^{(n)},\phi_2^{(n)}, 
          \psi^{(n)}|p^n_{X_1X_2},p^n_{K_1K_2})
        \geq 1-\tau.
	\end{align*} 
Hence we have the following theorem. 
\begin{theorem}
\label{th:SStConvTh}
For each $(\varepsilon, \delta) 
\in (0,\varepsilon_0] \times (0,1)$, we have
\begin{align}
\mathcal{R}(\varepsilon,\delta|p_{X_1X_2},p_{K_1K_2})
\subseteq  \mathcal{R}_{\mathrm {sw}}(p_{X_1X_2}). 
\notag
\end{align} 
\end{theorem}

In this paper we prove that 
for some $\varepsilon_0>0$, 
the set $\mathcal{R}_{\mathrm{key}}($ $p_{K_1K_2})$ 
serves as an outer bound of 
$\mathcal{R}(\varepsilon,\delta|p_{X_1X_2},$ $p_{K_1K_2})$ 
for $ (\varepsilon,\delta) \in (0,\varepsilon_0] \times (0,1)$.
As an immediate consequence of 
Proposition \ref{pro:ProOneA} part c), we have 
the following proposition.
\begin{proposition}
If $(R_1,R_2) \in 
\mathcal{R}(\varepsilon,\delta|p_{X_1X_2},p_{K_1K_2})$, then we
have that $\forall \gamma>0$, $\exists n_0(\gamma)$, 
$\forall n \geq n_0(\gamma)$, we have
\begin{align*}
& R_i \leq H(K_i)+\gamma +\frac{\varepsilon}{n}, i=1,2,
%\\
%& R_2 \leq H(K_2)+\gamma +\frac{\varepsilon}{n},
\\
& R_1+R_2 \leq H(K_1K_2)+\gamma +\frac{\varepsilon}{n}. 
\end{align*} 
\end{proposition}

From this proposition we have the following theorem.
\begin{theorem}
\label{th:SStConvThTwo}
For each $(\varepsilon, \delta) 
\in (0,\varepsilon_0] \times (0,1)$, we have
\begin{align}
\mathcal{R}(\varepsilon,\delta|p_{X_1X_2},p_{K_1K_2})
\subseteq  \mathcal{R}_{\mathrm {key}}(p_{K_1K_2}). 
\notag
\end{align} 
\end{theorem}

Combining Theorems 
\ref{th:directTh}, 
\ref{th:SStConvTh}, 
and \ref{th:SStConvThTwo}, we establish the following:
\begin{theorem}\label{th:StConvTh}
For each $(\varepsilon, \delta) 
\in (0,\varepsilon_0] \times (0,1)$, we have
\begin{align}
&      \mathcal{R}_{\mathrm{key}}(p_{K_1K_2})
  \cap \mathcal{R}_{\mathrm {sw}}(p_{X_1X_2})
\notag\\
&=\mathcal{R}(p_{X_1X_2},p_{K_1K_2})  
=\mathcal{R}(\varepsilon,\delta|
                      p_{X_1X_2},p_{K_1K_2}).
\notag
\end{align} 
\end{theorem}

\section{
Outline of the Proof of Theorem \ref{th:directTh} 
%Proposed Idea: Affine Encoders as Privacy Amplifier
}

%%%%%%%%%%%%%%%%%%%%%%%%%%%%%%%%%%%%%%%%%%%%%%%%%%%%%%%%%%%%%%%%% 
\begin{figure*}[t] 
%        \begin{center}	
	\centering 
	\includegraphics[width=0.70\textwidth]{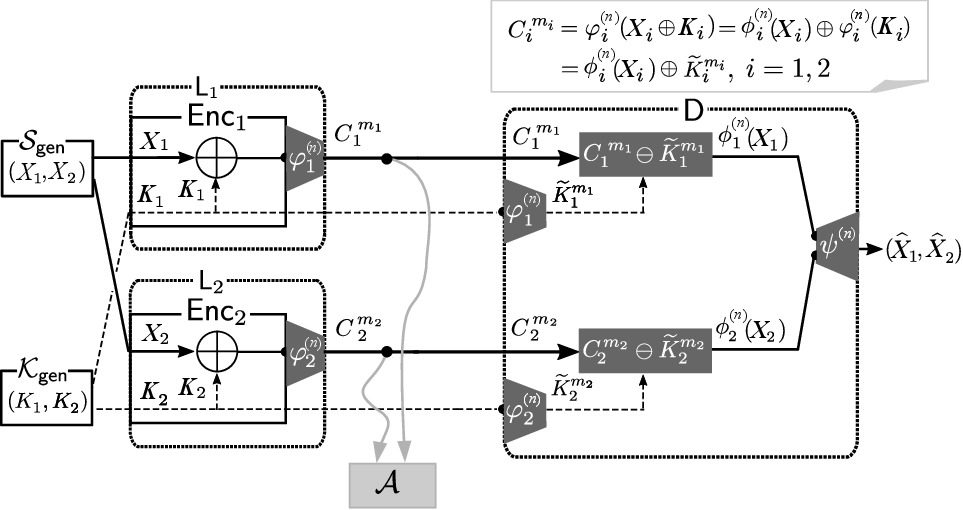}
	\caption{Our proposed solution: linear encoders 
	as privacy amplifiers.
\label{fig:solution}}%
%	\end{center}
\end{figure*}
%%%%%%%%%%%%%%%%%%%%%%%%%%%%%%%%%%%%%%%%%%%%%%%%%%%%%%%%%%%%%%%%%

%
%-------------------------------------------------------------

In this section we outline the proof of Theorem \ref{th:directTh}. 
Our construction of $(\Phi_1^{(n)},\Phi_2^{(n)},\Psi^{(n)})$ is 
the same as  that of Santoso and Oohama \cite{DBLP:conf/isit/SantosoO17}, 
\cite{santosoOhPEC:19} for the post encryption commpression scheme.    

Let $\phi^{(n)}:=(\phi_1^{(n)},\phi_2^{(n)})$ be 
a pair of linear mappings 
$\phi_1^{(n)}:\mathcal{X}_1^n\rightarrow \mathcal{X}_1^{m_1}$
and 
$\phi_2^{(n)}:\mathcal{X}_2^n\rightarrow \mathcal{X}_2^{m_2}$.
For each $i=1,2$, we define 
the mapping $\phi_i^{(n)} {\cal X}_i^n \to {\cal X}_i^{m_i}$
by 
\beq
\phi_i^{(n)}({\vcxi})
={\vcxi} A_i \mbox{ for }{\vcxi} \in {\cal X}_i^n,
\label{eq:homomorphica}
\eeq
where $A_i$ is a matrix with $n$ rows 
and $m_i$ columns. For each $i=1,2$, entries of $A_i$ are 
from ${\cal X}_i$. We fix $b_i^{m_i}\in \mathcal{X}_i^{m_i},$ 
$i=1,2$. For each $i=1,2$, define the mapping 
$\varphi^{(n)}_{i}: {\cal X}_i^n \to {\cal X}_i^{m_i}$
by 
\begin{align}
\varphi_{i}^{(n)}({\vcki}):=
\phi_i^{(n)}({\vcki})\oplus b_i^{m_i}
={\vcki}A_i\oplus b_i^{m_i}, 
%\mbox{ for }{\vcki} \in \mathcal{X}^n_i.
\label{eq:homomorphic}
\end{align}
for ${\vcki} \in \mathcal{X}^n_i$.
For each $i=1,2$, the mapping $\varphi_{i}^{(n)}$ 
is called the affine mapping induced by the linear mapping 
$\phi_{i}^{(n)}$ and constant vector 
$b_i^{m_i}$ $\in{\cal X}^{m_i}$.
For each $i=1,2$, define $\Phi_i^{(n)}$ by
$$
\Phi_i^{(n)}(\vck_i,\vcxi)=\varphi_i^{(n)}
({\vck}_i \oplus \vcxi). 
$$
By the definition (\ref{eq:homomorphic}) of 
$\varphi_i^{(n)}$, $i=1,2$, we have 
%those satisfy the following 
%affine structure: 
\begin{align}
&\Phi_i^{(n)}(\vck_i,\vcxi)=\varphi_i^{(n)}({\vcxi} \oplus {\vcki})
\notag\\
&= ({\vcxi} \oplus {\vcki})A_i\oplus b_i^{m_i}
={\vcxi} A_i\oplus({\vcki}A_i\oplus b_i^{m_i})
\notag\\
&=\phi_i^{(n)}({\vcxi})\oplus \varphi_i^{(n)}({\vcki}),
\mbox{ for } {\vcxi}, {\vcki} \in {\cal X}_i^n.
\label{eq:affine}
\end{align}
Set $\varphi^{(n)}:=(\varphi_1^{(n)},\varphi_2^{(n)})$. 
%%%%%%%%%%%%%%%%%%%%%%%%%%%%%%%%%%%%%%%%%%%%%%%%%%%%%%%%%%%%%%%%
%As a result from the linear structure 
%of $\phi_i^{(n)}$, $i=1,2$, $\phi_i^{(n)}$ $i=1,2$ satisfy 
%the following \emph{homomorphic} property:  
%\begin{align}
%	\phi_i^{(n)}(A_i^n\oplus B_i^n)=
%	\phi_i^{(n)}(A_i^n)\oplus \phi_i^{(n)}(B_i^n),\ i=1,2.
%\end{align}
%$$
%\varphi_{i}^{(n)}(x_i^n\Lambda_i):=
%\phi_i^{(n)}(x_i^n\Lambda_i)\oplus a_i^{m_i},
%\mbox{ for }x_i^n \in \mathcal{X}^n_i.
%$$
%%%%%%%%%%%%%%%%%%%%%%%%%%%%%%%%%%%%%%%%%%%%%%%%%%%%%%%%%%%%%%%%%
Next, let $\psi^{(n)}$ be the corresponding joint decoder 
for $\phi^{(n)}$ such that
$
\psi^{(n)}:\mathcal{X}_1^{m_1} 
\times \mathcal{X}_2^{m_2} \rightarrow 
\mathcal{X}_1^{n} \times \mathcal{X}_2^{n}.
$
Note that $\psi^{(n)}$ does 
not have a linear structure in general.
%For each $i=1,2$, we fix a vector $b_i^{m_i}\in {\cal X}^{m_i}$ 
%with $m_i$ columns. 
%---------------------------------------------------------
\subsubsection*{%
Description of Proposed procedure
}
%---------------------------------------------------------
We describe the procedure of our privacy amplified system as follows. 
%Our procedure only requires
%a pair of linear encoders
\begin{enumerate}
	\item\emph{Encoding of Ciphertexts:}
	First, we use $\varphi_1^{(n)}$ and $\varphi_2^{(n)}$
	to encode the ciphertexts $\rvcxone \oplus \rvckone$
	and $ \rvcxtwo \oplus \rvcktwo$.
	Let $C_i^{m_i}=\varphi_i^{(n)}(\rvcxi \oplus {\rvck}_i)$ 
        for $i=1,2$. 
%%%%%%%%%%%%%%%%%%%%%%%%%%%%%%%%%%%%%%%%%%%%%%%%%%%%%%%%%%%%%%%
%Then, instead of sending ${\rvccone}$ and ${\rvcctwo}$, 
%	we send $\tilde{C}_1^{m_1}$ and $\tilde{C}_2^{m_2}$
%	to public communication channel.
%%%%%%%%%%%%%%%%%%%%%%%%%%%%%%%%%%%%%%%%%%%%%%%%%%%%%%%%%%%%%%
        By the affine structure (\ref{eq:affine})
        of encoders we have that for each $i=1,2$,  
        \begin{align}
        %\widetilde
        &\Phi_i^{(n)}({\rvcki},{\rvcxi})=
        {C}_i^{m_i}=\varphi_i^{(n)}({\rvcxi}\oplus {\rvcki})
        \notag\\
        &=\phi_i^{(n)}({\rvcxi}) \oplus \varphi_i^{(n)}({\rvcki})
        % \notag\\
        %&
        =\widetilde{X}_i^{m_i} \oplus \widetilde{K}_i^{m_i},
        \label{eqn:aSdzx} 
        \end{align}
        where %we set 
        $\widetilde{X}_i^{m_i} \defeq \phi_i^{(n)}({\rvcxi}),
        \widetilde{K}_i^{m_i} \defeq \varphi_i^{(n)}({\rvcki}).$ 
        %for $i=1,2$.
	\item\emph{Decoding at Joint Sink Node $\D$:}
	First, using the pair of linear encoders
	$(\varphi_1^{(n)},\varphi_2^{(n)})$,
	$\D$ encodes the keys $({\rvckone},{\rvcktwo})$ 
         which are received through private channel into
	$(\widetilde{K}_1^{m_1},\widetilde{K}_2^{m_2})=$
	$(\varphi_1^{(n)}({\rvckone}),\varphi_2^{(n)}({\rvcktwo}))$.
	Receiving $(
	{C}_1^{m_1},
        {C}_2^{m_2})$ from public communication channel, $\D$ computes
	$\widetilde{X}_i^{m_i},i=1,2$ in the following way. From 
        (\ref{eqn:aSdzx}), we have that for each $i=1,2$, the decoder 
        $\D$ can obtain $\widetilde{X}_i^{m_i}=   \phi_i^{(n)}({\rvcxi})$
        by subtracting  $\widetilde{K}_i^{m_i}=\varphi_i^{(n)}({\rvcki})$
        from ${C}_i^{m_i}$. Finally, $\D$ outputs $(\hrvcxone, \hrvcxtwo)$
	by applying the joint decoder $\psi^{(n)}$ to
	$(\widetilde{X}_1^{m_1},\widetilde{X}_2^{m_2})$ as follows:
	\begin{align}
		(\hrvcxone, \hrvcxtwo)
		&=(\psi^{(n)}(\widetilde{X}_1^{m_1},
                  \widetilde{X}_2^{m_2})) 
         \notag\\
         &=(\psi^{(n)} (\phi_1^{(n)}({\rvcxone}),
                              \phi_2^{(n)}({\rvcxtwo})). 
                 \label{eq:source_estimation}
	\end{align}
We summarize the above argument. For $(\rvckone,\rvcktwo)$  
and $(C_1^{m_1},C_2^{m_2})$, define $\Psi^{(n)}$ by 
\begin{align*}
&\Psi^{(n)}(
\rvckone,
\rvcktwo,
C_1^{m_1},
C_2^{m_2})=\Psi^{(n)}_{\lrvckone,\lrvcktwo}(
C_1^{m_1},
C_2^{m_2}) 
\\
& \defeq \psi^{(n)}(
{C}_1^{m_1}\ominus \widetilde{K}_1^{m_1},
{C}_2^{m_2}\ominus \widetilde{K}_2^{m_2})
\\
&
=\psi^{(n)}(\widetilde{X}_1^{m_1},\widetilde{X}_2^{m_2}).
\end{align*}
By the above definition and 
$C_i^{m_i}=\Phi^{(n)}_{i,{\lrvck}_i}(\rvcxi),i=1,2,$ we have 
\begin{align*}
&\Psi^{(n)}_{\lrvckone,\lrvcktwo}(
\Phi^{(n)}_{1,\lrvckone}(\rvcxone),
\Phi^{(n)}_{2,\lrvcktwo}(\rvcxtwo))
\\
&=\psi^{(n)}(\widetilde{X}_1^{m_1},\widetilde{X}_2^{m_2})
%\\
%&
=\psi^{(n)}(\phi_1^{(n)}(\rvcxone),
              \phi_2^{(n)}(\rvcxtwo)).
\end{align*}
Hence we have the condition which 
$(\Phi^{(n)}_{1}, \Phi^{(n)}_{2},\Psi^{(n)})$
must satisfy.
\end{enumerate}
In this paper, we use the \emph{minimum entropy decoder} 
for our joint decoder $\psi^{(n)}$. \\
\underline{\it Minimum Entropy Decoder:} \ 
For 
$\phi_i^{(n)}(\vcxi)=\widetilde{x}_i^{m_i},i=1,2$, 
%we define the joint decoder function 
$
\psi^{(n)}: {\cal X}_1^{m_1}\times {\cal X}_2^{m_2}
\to {\cal X}_1^{n} \times {\cal X}_2^{n}
$ is defined as follows:
\begin{align*}
&\psi^{(n)}(\widetilde{x}_1^{m_1},\widetilde{x}_2^{m_2})
\\
&:=\left\{\begin{array}{cl}
({\hvcxone}, {\hvcxtwo})
   &\mbox{if } \phi_i^{(n)}({\hvcxi})=\widetilde{x}_i^{m_i},i=1,2, \\
   %&\mbox{\ \ \ } \phi_2^{(n)}({\hvcxtwo}{})=\widetilde{x}_2^{m_2},\\
   &\mbox{and }H({\hvcxone} {\hvcxtwo})
    <H( {\cvcxone} {\cvcxtwo})\\
   &\mbox{for all }({\cvcxone},{\cvcxtwo})\mbox{ such that }\\
   & \:\phi_i^{(n)}({\cvcxi})=\widetilde{x}_i^{m_i},i=1,2,\\
   %& \:\phi_2^{(n)}({\cvcxtwo}{})=\widetilde{x}_2^{m_2},\\
   & \mbox{and } 
    \:({\cvcxone},{\cvcxtwo})
    \neq ({\hvcxone},{\hvcxtwo}),
\vspace{0.2cm}\\
\mbox{arbitrary}
    & \mbox{if there is no such } %\\
    %& 
({\hvcxone},{\hvcxtwo})
     \in{\cal X}_1^{n}\times{\cal X}_2^{n}.
\end{array}
\right.
\end{align*}
Our privacy amplified system described above is illustrated 
in Fig. \ref{fig:solution}.

\noindent
\underline{\it Evaluations of the reliablility and security:} \ 
On the error probability 
$p_{\rm e}$ of decoding we have
\begin{align*}
&p_{\rm e}=\Pr[\Psi^{(n)}(\rvckone, \rvcktwo,
  \phi_1^{(n)}({\rvckone, \rvcxone}),
  \phi_2^{(n)}({\rvcktwo, \rvcxtwo}))
\\
&\qquad \neq ({\rvcxone},{\rvcxtwo}) ]
\\
&=\Pr[\Psi^{(n)}_{\lrvckone,\lrvcktwo}
(\Phi_{1,\lrvckone}^{(n)}(\rvcxone),
 \Phi_{2,\lrvcktwo}^{(n)}(\rvcxtwo))
\neq ({\rvcxone},{\rvcxtwo}) ]
\\
&=\Pr[\psi^{(n)}
(\phi_{1}^{(n)}(\rvcxone),
 \phi_{2}^{(n)}(\rvcxtwo))
\neq           (\rvcxone,
                \rvcxtwo)].
%\\
%&=\Pr[(\rvcxone,\rvcxtwo)\notin {\cal D}^{(n)})].
\end{align*}
%\underline{\it An Upper Bound of 
%$\Delta_{{\rm max-MI}}^{(n)}(\Phi_1^{(n)},\Phi_2^{(n)},
%         \Psi^{(n)}|{p}_{K_1K_2}^n)$:} 
We have the following upper bound of 
%$\Delta_{{\rm max-MI}}^{(n)}(\Phi^{(n)},
%         \varphi_{\cal A}^{(n)}|{p}_{KZ}^n)$. 
$\Delta_{{\rm max-MI}}^{(n)}(\Phi_1^{(n)},\Phi_2^{(n)},$ 
$ \Psi^{(n)}|{p}_{K_1K_2}^n)$.
\begin{lemma}\label{lem:LemB} 
For the proposed construction of 
$(\Phi^{(n)}_1,\Phi^{(n)}_2,$ $\Psi^{(n)})$, 
we have 
\begin{align*}
& \Delta_{{\rm max-MI}}^{(n)}(\Phi_1^{(n)},\Phi_2^{(n)},
         \Psi^{(n)}|{p}_{K_1K_2}^n)
\\
&\leq \overline{\Delta}_{{\rm max-MI}}^{(n)}
(\Phi_1^{(n)},\Phi_2^{(n)}|{p}_{K_1K_2}^n)
\\
&\leq  m_1 \log |{\cal X}_1| 
 +m_2 \log |{\cal X}_2|
 -H(\widetilde{K}_1^{m_1}\widetilde{K}_2^{m_2})
\\
&=D(p_{\widetilde{K}_1^{m_1}
      \widetilde{K}_2^{m_2}} \|
    p_{{U}_1^{m_1}
       {U}_2^{m_2}}). 
\end{align*}
Here $p_{{U}_1^{m_1}{U}_2^{m_2}}$ is the uniform distribution 
over ${\cal X}_1^{m_1}\times {\cal X}_2^{m_2}$.
\end{lemma}

\begin{IEEEproof}
By Proposition \ref{pro:ProOneA} part a), it sufficies to prove 
the upper bound of   
$
\overline{\Delta}_{{\rm max-MI}}^{(n)}
(\Phi_1^{(n)},\Phi_2^{(n)}|{p}_{K_1K_2}^n).
$
For the proposed construction of $(\Phi_1^{(n)},\Phi_2^{(n)})$, 
we have 
\begin{align}
\overline{C}_i^{m_i}=\wt{K}_i^{m_i}
\oplus \phi_i^{(n)}(\overline{\rvcx}_i), i=1,2.
\label{eqn:Sdddz}
\end{align}
Then we have the following chain of inequalities:
\begin{align}
& I(\overline{C}_1^{m_1}\overline{C}_1^{m_1}; 
   \overline{\rvcx}_1 \overline{\rvcx}_2)
\notag\\
&=  H(\overline{C}_1^{m_1}\overline{C}_2^{m_2})
   -H(\overline{C}_1^{m_1}\overline{C}_2^{m_2}
       |\overline{\rvcx}_1\overline{\rvcx}_2)
\notag\\
&\leq m_1 \log |{\cal X}_1| 
  +m_2 \log |{\cal X}_2|
  -H(\overline{C}_1^{m_1}\overline{C}_2^{m_2}|
 \overline{\rvcx}_1 
 \overline{\rvcx}_2)
\notag\\
&\MEq{a} m_1 \log |{\cal X}_1| 
  +m_2 \log |{\cal X}_2|
\notag\\
&\quad 
-H(\widetilde{K}_1^{m_1}\oplus 
\phi_1^{(n)}(\overline{\rvcx}_1),     
\widetilde{K}_2^{m_2}\oplus     
\phi_2^{(n)}(\overline{\rvcx}_2)
    |\overline{\rvcx}_1\overline{\rvcx}_2)
\notag\\
&= m_1 \log |{\cal X}_1| 
  +m_2 \log |{\cal X}_2|
 -H(\widetilde{K}_1^{m_1}\widetilde{K}_2^{m_2}
    |\overline{\rvcx}_1\overline{\rvcx}_2)
\notag\\
&\MEq{b} m_1 \log |{\cal X}_1| 
  +m_2 \log |{\cal X}_2|
 -H(\widetilde{K}_1^{m_1}\widetilde{K}_2^{m_2})
\notag\\
&=D(p_{\widetilde{K}_1^{m_1}
      \widetilde{K}_2^{m_2}} \|
    p_{{U}_1^{m_1}
       {U}_2^{m_2}}). 
\label{eqn:AsDDss}
\end{align}
Step (a) follows from (\ref{eqn:Sdddz}).
Step (b) follows from 
$(\overline{\rvcx}_1,
  \overline{\rvcx}_2)$ 
$\perp (\rvckone,\rvcktwo)$.
Since (\ref{eqn:AsDDss}) holds for any 
$(\overline{\rvcx}_1,\overline{\rvcx}_2)$, 
we have the upper bound of 
$
\overline{\Delta}_{{\rm max-MI}}^{(n)}
(\Phi_1^{(n)},\Phi_2^{(n)}|{p}_{K_1K_2}^n)
$ in Lemma \ref{lem:LemB}. 
\end{IEEEproof}
According to Santoso and Oohama \cite{santosoOhPEC:19}, 
$\exists \{(\Phi_1^{(n)},\Phi_2^{(n)},$ $\Psi^{(n)})\}_{n \geq 1}$
such that for any $(p_{X_1X_2},p_{K_1K_2})$ satisfying  
\begin{align*}
&\left(\frac{m_1}{n}\log |{\cal X}_1|, 
 \frac{m_2}{n}\log |{\cal X}_2|\right)
\\
& \in 
\mathcal{R}_{\mathrm{key}}(p_{K_1K_2})
\cap 
\mathcal{R}_{\mathrm{sw}}(p_{X_1X_2}),
\end{align*}
the two quantities 
\begin{align*}
& \Pr[\psi^{(n)}
(\phi_{1}^{(n)}(\rvcxone),
 \phi_{2}^{(n)}(\rvcxtwo))
\neq           (\rvcxone,
                \rvcxtwo)]
\mbox{ and }
\\
& D(p_{\widetilde{K}_1^{m_1}
      \widetilde{K}_2^{m_2}} \|
    p_{{U}_1^{m_1}
       {U}_2^{m_2}})
\end{align*}
decay exponentially as $n$ tends to infinity. Hence we 
have Theorem \ref{th:directTh}.   

%\begin{remark}
%In \cite{DBLP:conf/isit/SantosoO17}, \cite{santosoOhPEC:19}, 
%on the surface, Santoso and Oohama stated that their
%distributed encryption scheme is secure under the security metric
%based on $\Delta_{\rm MI}^{(n)}$. However, if one looks underneath a
%little bit, one can easily discover that actually Santoso and Oohama
%proved the security of their scheme based on a metric that is more
%strict than $\Delta_{\rm MI}^{(n)}$. And one can easily see that the
%more strict security metric they used is equal to 
%our new proposed metric $\Delta^{(n)}$. Thus, 
%$\Delta^{(n)}$ is not a mere theoretical concept, but an achievable 
%security requirement which appears naturally in the secure analysis 
%of the communication system treated here. 
%\end{remark}

\appendix

\ProofPrOnDecSet
\ProofPrOne
\ProofLemOne
%\ProofKeyLm
\ProofProOneA

\bibliographystyle{IEEEtran}

\bibliography{Isita2022}
%\bibliography{RefEdByOh}

\end{document}